\documentclass[portrait,iop,revtex4]{emulateapj}
\slugcomment{Accepted for publication in {\it The Astrophysical Journal} September 12, 2014}
\shortauthors{M. Latour et al.}
\shorttitle{Helium-Carbon Correlation on the EHB in $\omega$ Centauri}
\newcommand{\gta}{\lower 0.5ex\hbox{$ \buildrel>\over\sim\ $}}
\newcommand{\lta}{\lower 0.5ex\hbox{$ \buildrel<\over\sim\ $}}
\newcommand{\nhe} {$N$({\rm He})/$N$({\rm H})}
\newcommand{\nc} {$N$({\rm C})/$N$({\rm H})}
\newcommand{\teff}{$T_{\rm eff}$}
\newcommand{\Msun}{$M_{\rm \odot}$}

\usepackage{natbib}
\usepackage{graphicx}
\usepackage{amsmath}
\begin{document}

\title{A Helium-Carbon Correlation on the Extreme Horizontal Branch in $\omega$ Centauri\footnote{Based on observations collected at the
    European Organisation for Astronomical Research in the Southern
  Hemisphere, Chile (proposal ID 386.D-0669 and 091.D-0791)}}

\author{M. Latour\altaffilmark{1,2}, S. K. Randall\altaffilmark{3},
   G. Fontaine\altaffilmark{1}, G. Bono\altaffilmark{4,5}, A. Calamida\altaffilmark{6} and P. Brassard\altaffilmark{1}}

\altaffiltext{1}{D\'epartement de Physique, Universit\'e
  de Montr\'eal, Succ. Centre-Ville, C.P. 6128, Montr\'eal, QC H3C 3J7,
  Canada} 
\altaffiltext{2} {Dr. Karl Remeis-Observatory \& ECAP, Astronomical Institute,
Friedrich-Alexander University Erlangen-Nuremberg, Sternwartstr. 7, 96049 Bamberg, Germany} 
\altaffiltext{3}{ESO, Karl-Schwarzschild-Str. 2, 85748 Garching bei
  M\"{u}nchen, 
  Germany} 
\altaffiltext{4}
 {Istituto Nazionale de Astrofisica, Osservatorio Astronomico di Roma, via Frascati 33, 00040 Monte Porzio Catone, Italy}

\altaffiltext{5}{
 Universit\`{a} di Roma ``Tor Vergata'', Department of Physics, via della Ricerca Scientifica 1, 00133 Rome, Italy}

\altaffiltext{6}{Space Telescope Science Institute, 3700 San Martin
  Drive, Baltimore, MD 21218}

\begin{abstract}
Taking advantage of a recent FORS2/VLT spectroscopic sample of Extreme Horizontal 
Branch (EHB) stars in $\omega$ Cen, we isolate 38 spectra well suited for detailed
atmospheric studies and determine their fundamental parameters (\teff,
log $g$, and log \nhe) using NLTE, metal
line-blanketed models. We find that our targets can be divided into three
groups: 6 stars are hot (\teff~\gta 45,000 K) 
H-rich subdwarf O stars, 7 stars are typical H-rich sdB stars (\teff~\lta
35,000 K), and the remaining 25 targets at intermediate effective
temperatures are He-rich (log \nhe~\gta $-$1.0) subdwarfs. Surprisingly
and quite interestingly, these He-rich hot subdwarfs in $\omega$ Cen
cluster in a narrow temperature range ($\sim$35,000 K to
$\sim$40,000 K). We additionally measure the
atmospheric carbon abundance and find a most interesting positive 
correlation between the carbon and helium atmospheric abundances. 
This correlation certainly bears the signature of diffusion 
processes - most likely gravitational settling impeded by stellar winds 
or internal turbulence - but also constrains possible formation scenarios
proposed for EHB stars in $\omega$ Cen. For the He-rich objects in particular,
the clear link between helium and carbon enhancement points towards a 
late hot flasher evolutionary history.

\end{abstract}


\section{ASTROPHYSICAL CONTEXT}

Globular clusters (GCs) are ideal laboratories for constraining
the evolutionary properties of low-mass stars
and investigating the formation and kinematic evolution of
low--mass stellar systems \citep{dic13,zoc12}.
The key advantages in dealing with cluster stellar populations are
manifold: cluster stars have the same age and the same iron
abundance. Moreover, they are located at the same distance and 
are typically characterized by the same reddening.
However, several of the above assumptions concerning cluster simple
stellar populations have been challenged by both spectroscopic and
photometric evidence. It has been shown that most GCs host at least two
generations of stars differing mainly in
light element abundance \citep{car10}. The second stellar generation is thought to be formed from material polluted by the first generation but it is still not clear exactly how this pollution occurred.  Possible polluters might be asymptotic giant branch (AGB) stars during their thermal pulse phase  \citep{ven01,gra04},or fast rotating massive stars \citep{mae06,dec07}. In any case, the pollution and the subsequent formation of a second generation of stars happen soon enough in the life of the cluster (in $\sim$10$^7$-10$^8$ yr) that the age spread is usually not detectable at the level of the main sequence turn off \citep{gra04}.

In the specific case of $\omega$ Cen, there are at least three separate stellar populations with a large undisputed spread in iron abundance (more than 1 dex) \citep{vil07,cal09,joh10}. 
$\omega$ Cen also has a double MS \citep{and97,bed04}, the bluer sequence being composed of more metal-rich stars that are also believed to be helium enhanced (Y $\sim$0.4; \citealt{pio05}).
The presence of stellar populations with different chemical composition has repercussions on the characteristics of the evolved stellar evolutionary phases,
such as the helium burning horizontal branch (HB) and its blue extension, the extreme horizontal branch (EHB)\footnote{ 
The EHB stars are hot (\teff\ $\gta$20,000 K) and compact core helium-burning objects. Their high temperature is caused by their very thin hydrogen envelope (M $\lta$0.02  $M_{\odot}$) that is not massive enough to sustain significant hydrogen shell burning.}.

Intriguingly, the occurrence of very blue (thus hot) HB stars in $\omega$ Cen (as well as in other massive globular clusters with complex populations) cannot be explained by canonical evolution \citep{dcruz96}. Instead, two competing non-canonical scenarios have been proposed. 

In the first scenario, the blue tail of the HB is explained by the presence of a He-enhanced second generation of stars. These stars leave the main sequence with a lower core mass at a given globular cluster age, resulting
in a higher temperature when they reach the helium burning phase on
the HB, thus qualitatively explaining its bluer morphology
(\citealt{dan02,bus07}). A key advantage of this scenario is that it can explain the EHB population assuming a standard  mass-loss efficiency along the RGB branch.  
In contrast, the second evolutionary scenario predicts that some stars experience a
helium core flash after having evolved away from the RGB.
This ``hot flasher'' scenario has been modeled by several groups
(e.g., \citealt{cas93}; \citealt{dcruz96}; \citealt{bro01}) and has indeed been shown to 
 produce helium burning stars that settle at the very hot end of the EHB.  Different ``flavors'' of 
hot flashers may occur depending on the evolutionary stage of the star at the time of the helium flash.  
If the ignition of helium happens before the star reaches the white dwarf
cooling curve, the hydrogen-burning shell forms a barrier that prevents
the inner convection zone from reaching the envelope of the star. This
situation is often referred to as an ``early hot flash'' and results in a
hot EHB star with a H-rich atmosphere. If on the other hand the
flash does not occur until the star has settled onto the white dwarf cooling
sequence, the reduced entropy of the much weaker hydrogen-burning shell allows the convection
zone to extend out to the surface, mixing the helium and carbon
from the core with the hydrogen present in the atmosphere. 
Depending on the degree of hydrogen shell burning at the time of the flash, the mixing efficiency may vary. In the most common ``late hot flasher'' case, the deep mixing is likely to burn most of the hydrogen carried into the
interior and the resulting star will thus arrive at the blue end of the EHB with an
atmosphere dominated by helium. The surface composition predicted from models of late
flashers is around 95 to 96 \% helium by mass and 3 to 4 \% carbon
(\citealt{bro01}; \citealt{cass03}). Evolutionary paths for different
types of hot flashers can be found in Figure 4 of \citet{bro01}. Note that in all these cases
 the star must have lost a large amount of its hydrogen envelope on the RGB via some mechanism.

The hot flasher scenario can also be invoked to explain EHB stars 
(spectral types sdO and sdB) among the Galactic field population, in particular the He-rich sdBs
 (\citealt{lanz04}; \citealt{mil08}). 
 He-poor sdB stars in the field can be modeled in terms of a canonical evolutionary scenario,
 where the He-flash occurs at the tip of the RGB \citep{dor93}. The necessary mass loss can also be 
explained in terms of binary interactions such as Roche lobe overflow and common envelope evolution, 
mergers involving at least one He white dwarf, or even planet ingestion (see \citealt{heb09} for a review of formation mechanisms). Complications appear with the helium-enriched
subdwarfs, whose existence cannot be explained by canonical evolution. In these cases,
alternate scenarios, such as the late helium flash
discussed above or the merger of two white dwarfs \citep{saio00}, are invoked. 
However, according to these scenarios, He-rich hot subdwarfs should
quickly settle on the zero age helium main sequence (ZAHEMS), which 
does not appear to be the case from observations \citep{stro07,heb06}. Understanding the formation of 
EHB stars, both in the field and in globular clusters, therefore remains a challenge.

In this context, it remains essential to characterize as many hot
subdwarfs as possible. Globular clusters EHB stars in particular have been studied less
 than their field counterparts due to the obvious observational difficulties. 
However, for $\omega$ Cen there have been several surveys of HB and EHB stars aimed 
at gaining insight on the formation
mechanism and evolutionary status of these objects (see, e.g.,
\citealt{moe02}, \citeyear{moe07}, \citeyear{moe11}, and
\citealt{moni2012}). 
These studies combined spectroscopic observations
and model atmosphere techniques to derive atmospheric parameters for
several hot subdwarf stars in that cluster. Quite interestingly, \citet{moe11}
found preliminary evidence for a correlation between carbon and helium enhancement. 
This would point toward the hot flasher scenario as the origin of helium-enriched stars, 
but does not rule out the possibility of the He-enhanced scenario also playing a role \citep{cass09}. 
Shedding light on the EHB evolution in globular clusters is indeed of great importance, 
not only for the understanding of the late evolutionary stages of low-mass stars, but also for the interpretation of the multiple populations observed in some GCs.

Another development concerning $\omega$ Cen and its population of hot
subdwarfs has been the recent discovery of short-period EHB pulsators as
reported by \citet{ran09,ran11}. Contrary to initial
expectations, these variables turned out not to be the analogs of field
sdB pulsators discovered almost two decades ago \citep{kil97}, but
members of a new family of H-rich sdO pulsating stars with effective 
temperatures clustering around 50,000 K. Interestingly, 
and despite extensive searches \citep{john13}, no field counterparts to
those sdO pulsators have been found\footnote{Pulsating hot subdwarfs
have also been discovered quite recently in the globular cluster NGC
2808 as reported by \citet{bro13} but fundamental parameters still have to be determined for some of them.}. In an effort to map this new $\omega$ Cen instability strip in
the log $g$-\teff~diagram, FORS spectroscopy was obtained
at the VLT for a sample of 60 EHB star candidates. 
Preliminary results for 19 stars were reported in \citet{ran13} and the
complete study will be presented elsewhere (Randall et al. 2014, in
preparation). For the purposes of the present study, we selected 38 
 stars whose spectra showed no signs of pollution from nearby stars for
 more detailed spectroscopic modeling. During the course of this we 
found that the majority of the sample have He-rich atmospheres and, 
moreover, that carbon features can be seen in 25 of them, thus opening 
up the possibility of investigating the relationship between helium 
richness and carbon abundance in hot subdwarfs in a globular cluster 
environment. In light of the currently raging debate on EHB star evolution 
we felt that this would be a most worthwhile endeavour to pursue. In Section 2 we thus present our
observational material, followed by the resulting fundamental parameters 
and carbon abundances measured for our stars in Section 3. Finally, in Section 4
we discuss our results and compare them to similar results for field star samples
 and theoretical predictions. 

\section{OBSERVATIONAL MATERIAL}

\begin{figure*}[p]
\includegraphics[angle=270,scale=0.6]{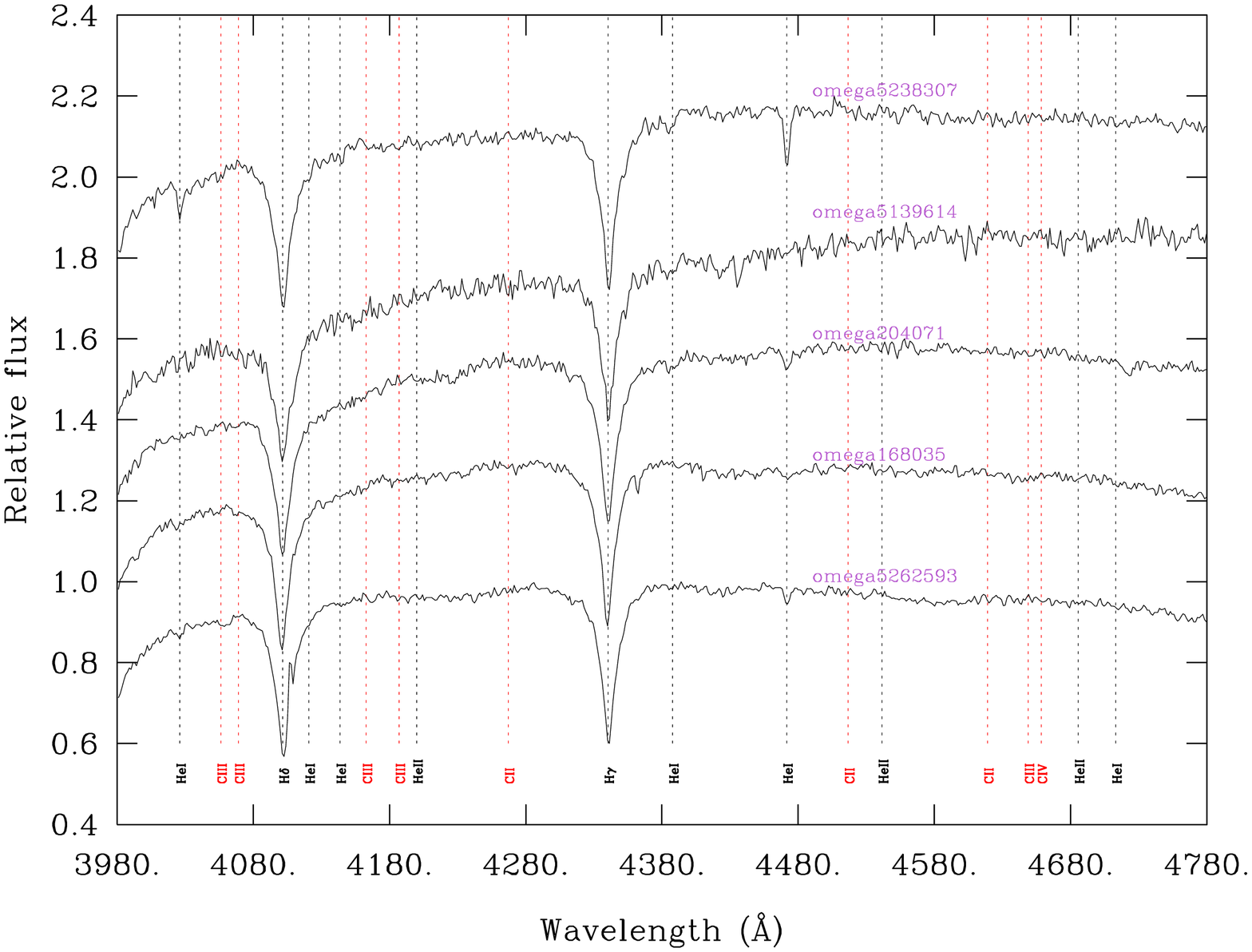}
\includegraphics[angle=270,scale=0.6]{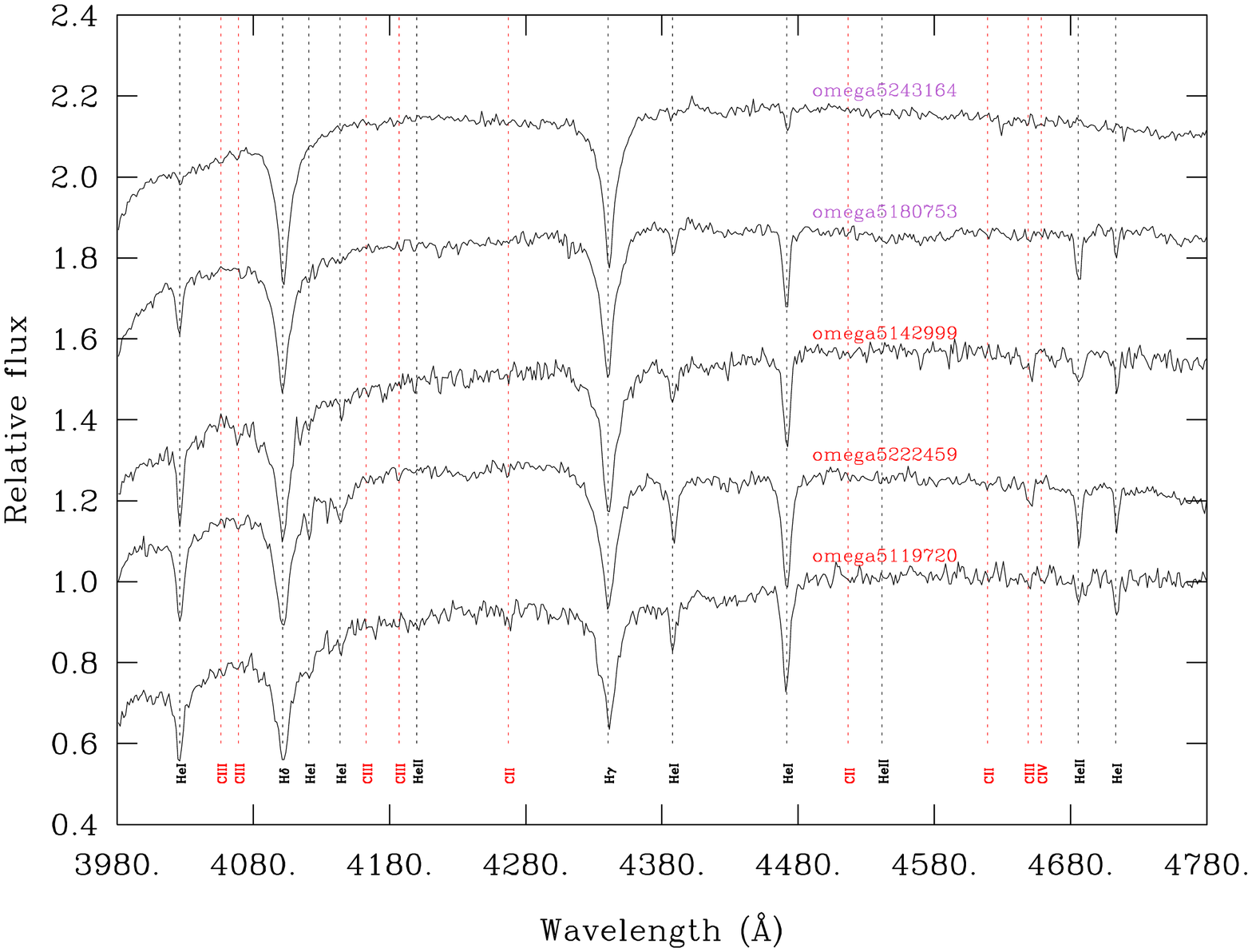}
 \caption{Spectra of the 38 stars that make up our
sample of EHB stars in $\omega$ Cen. The spectra are displayed over the 
3980$-$4780 \AA~range, which includes most of the interesting carbon
features. The locations of the relevant hydrogen and helium lines are also
indicated. The spectra are not flux calibrated, but they have been corrected for a radial velocity shift of 232.2 km s$^{-1}$, the accepted value for $\omega$ Cen \citep{har96}. The spectra are presented in the same order as in Table 1,
and the color used to indicate their names makes it easy to associate
them with their respective group (see Section 3). Panels a and b. 
}
 \label{spec}
\end{figure*}

\addtocounter{figure}{-1}
\begin{figure*}[p]
\includegraphics[angle=270,scale=0.6]{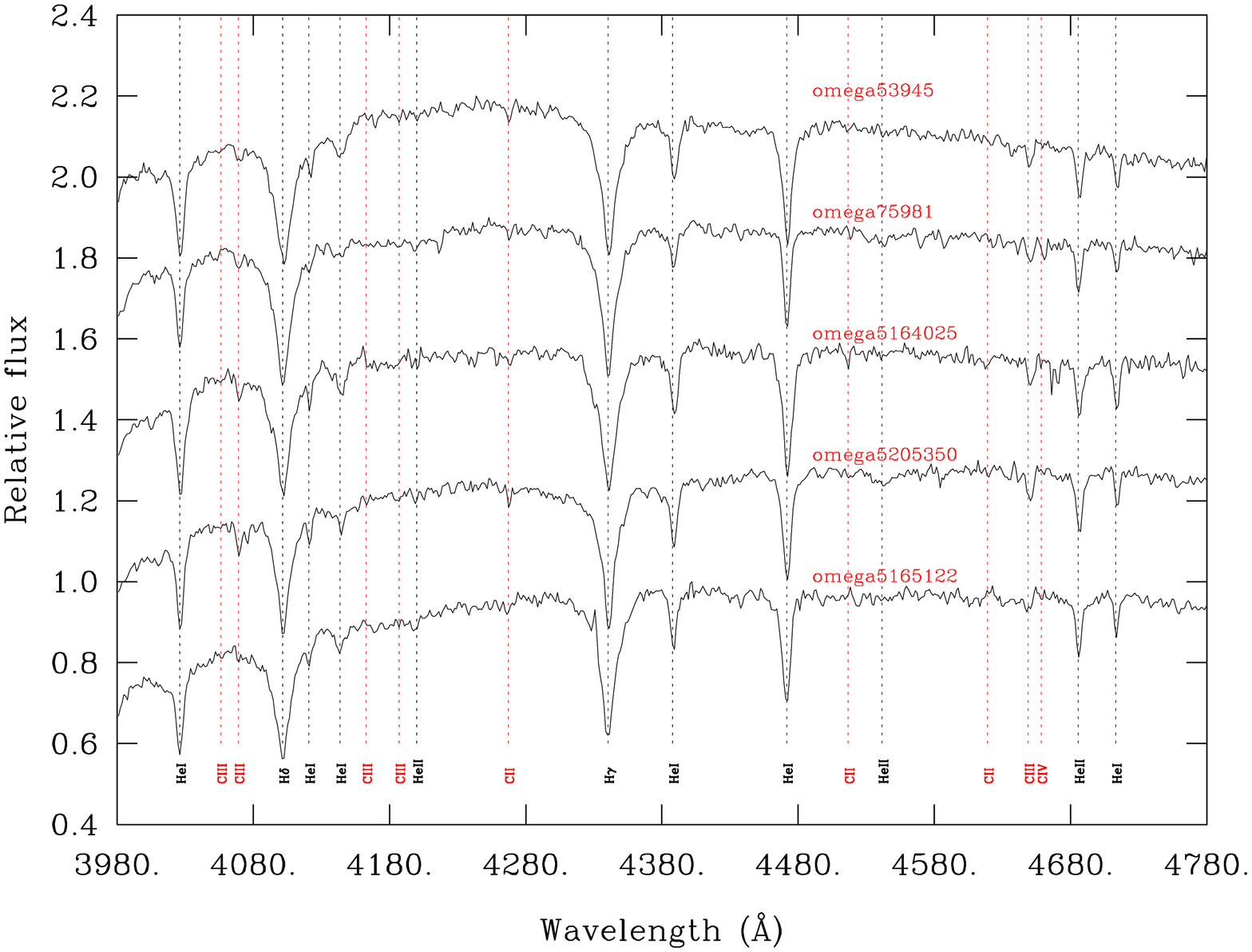}
\includegraphics[angle=270,scale=0.6]{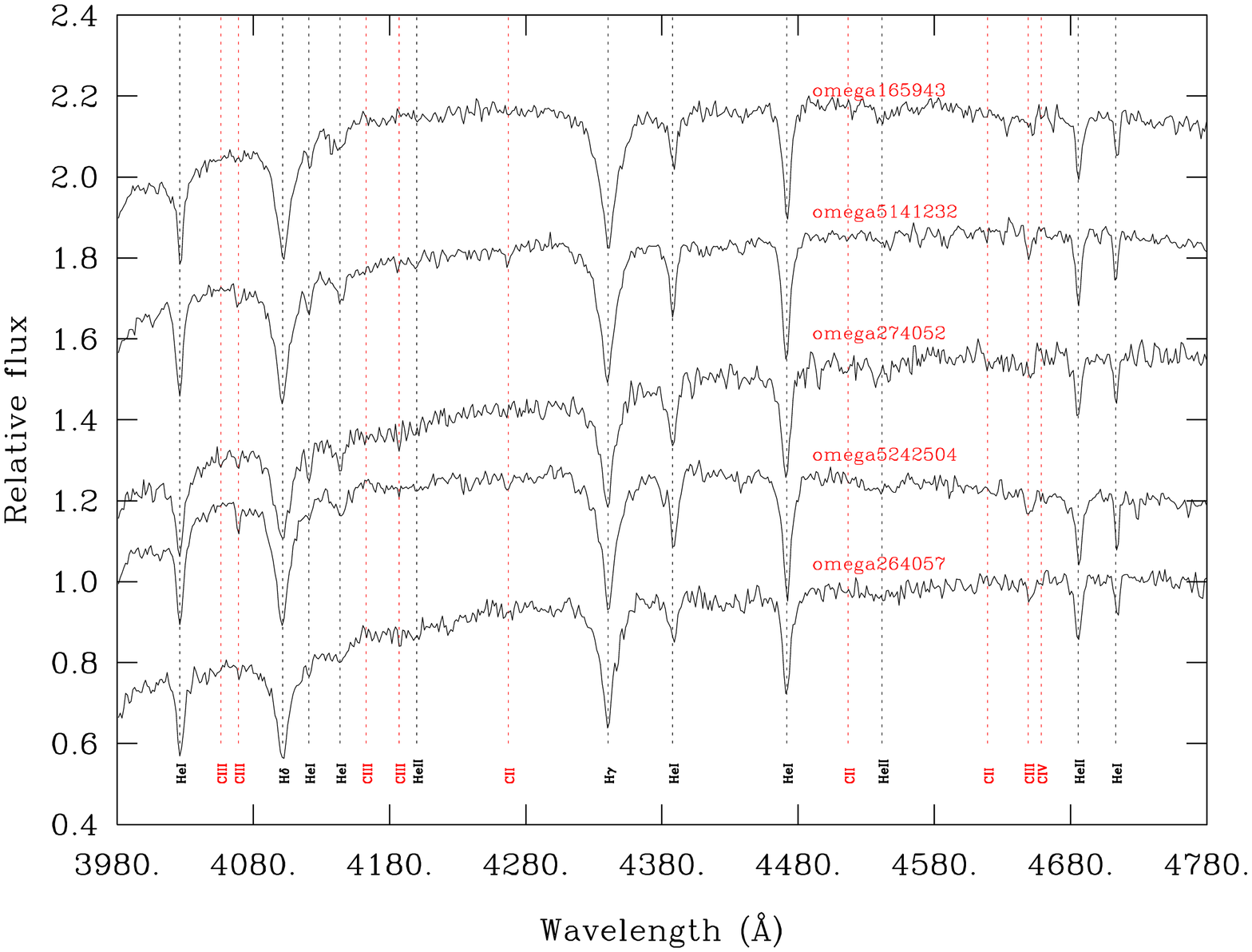}
 \caption{Continued. Panels c and d. 
}
\end{figure*}

\addtocounter{figure}{-1}
\begin{figure*}[p]
\includegraphics[angle=270,scale=0.6]{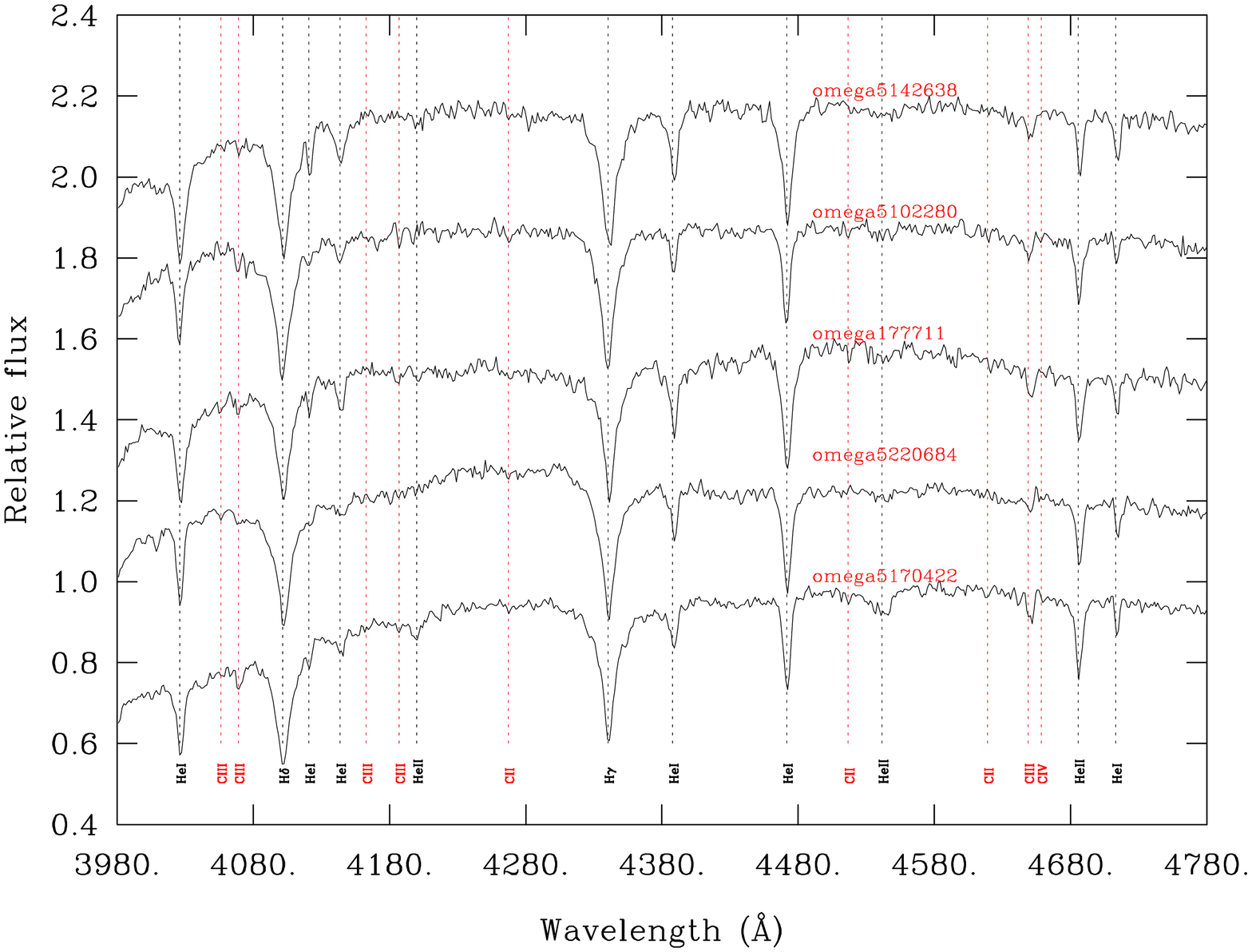}
\includegraphics[angle=270,scale=0.6]{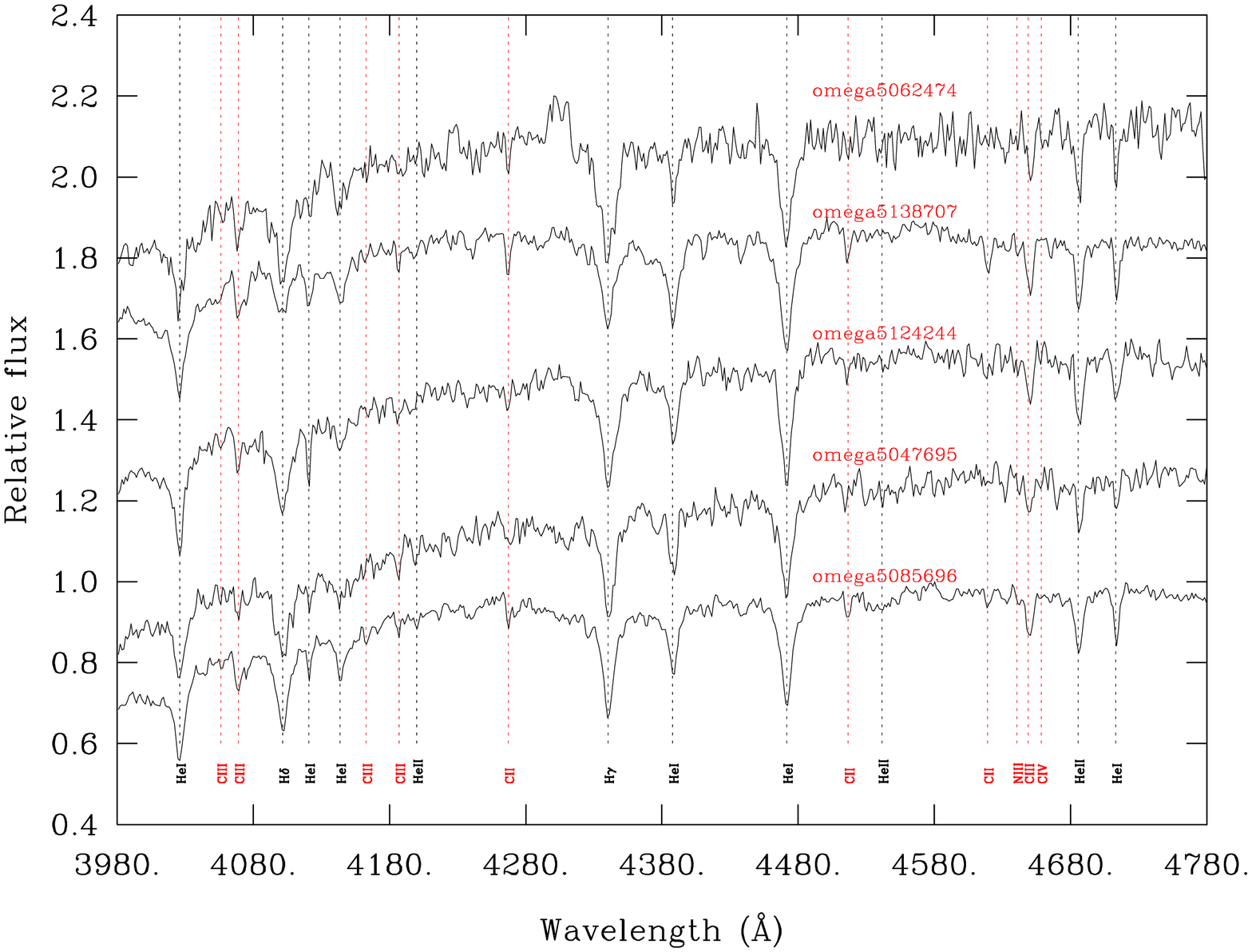}
 \caption{Continued. Panels e and f. 
}
\end{figure*}

\addtocounter{figure}{-1}
\begin{figure*}[p]
\includegraphics[angle=270,scale=0.6]{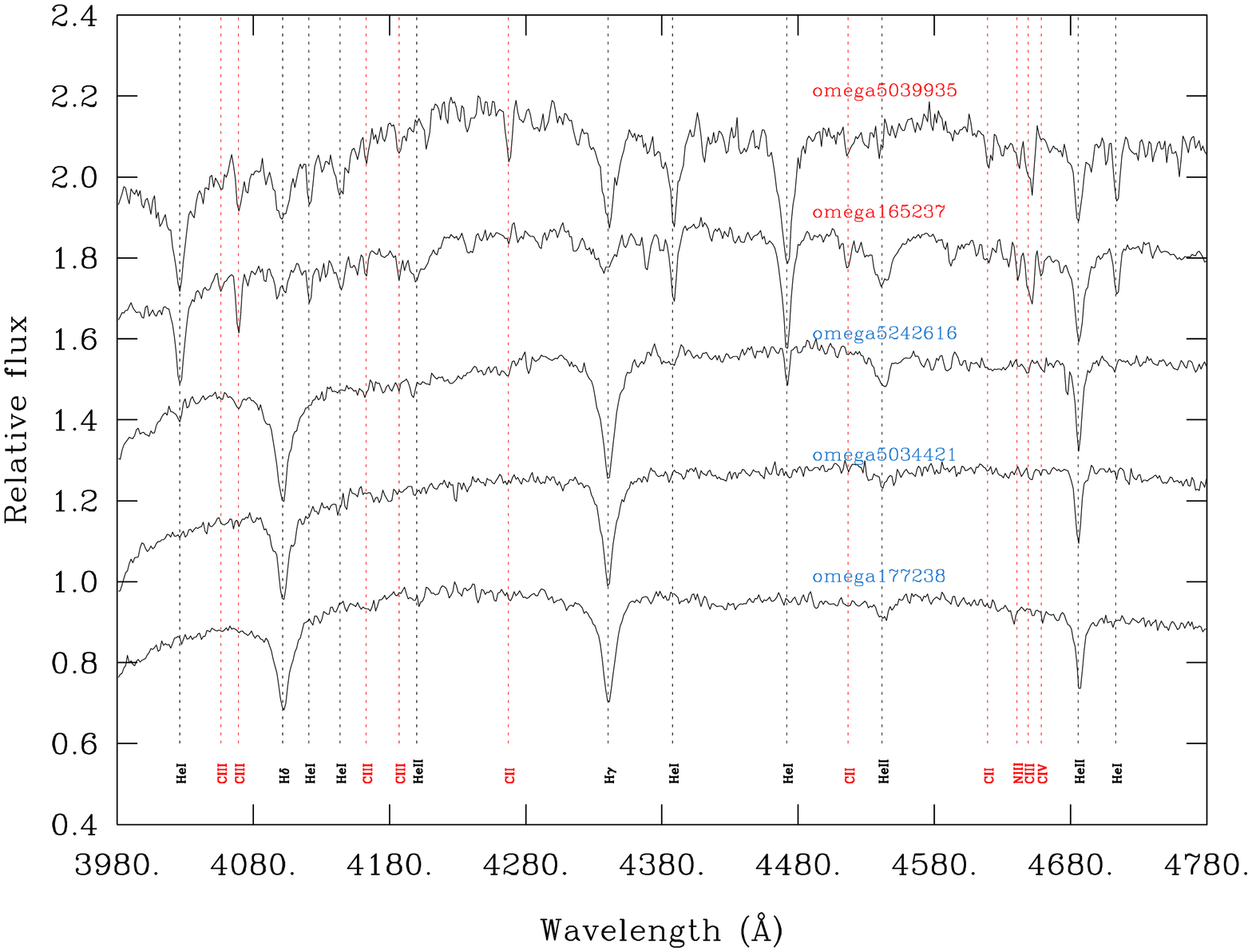}
\includegraphics[angle=270,scale=0.6]{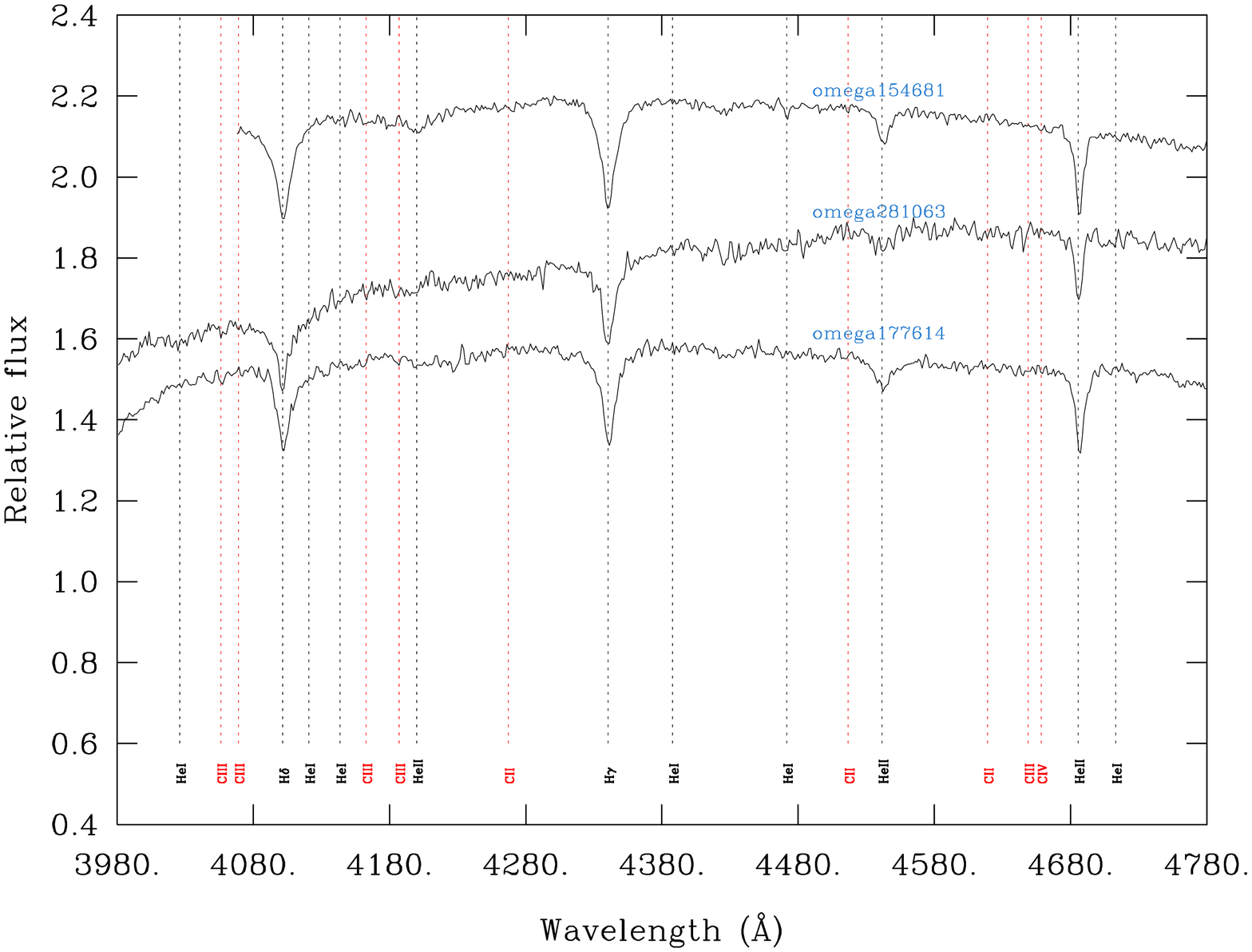}
 \caption{Continued. Panels g and h. 
}
\end{figure*}

The spectra used here were taken from the initial sample of 60 $\omega$
Cen EHB star candidates as described in detail by Randall et al. 2014 (in
preparation). The spectroscopic targets had been selected as EHB star candidates based
on their brightness and colour in the $\omega$ Cen WFI/ACS catalogue
\citep{cast07}. Since the primary aim of the observations was to map 
the sdO instability strip, the color cut favors the hotter part of the EHB domain
(\teff $\gta$ 30,000 K). For the present study we specifically excluded spectra 
that were too polluted by nearby stars to derive reliable atmospheric parameters, leaving us with 38
uncontaminated EHB star spectra to base our analysis upon.
Thus the stars in our sample feature the expected continuum slope -- no increase from blue to red -- and show no signs of the following pollution indicators: G-band, Mg~\textsc{i} triplet (5167, 5172, and 5183 \AA), Na~\textsc{i} doublet (5890 and 5896 \AA), and Ca~\textsc{ii} K line. Of course some of these lines originating from the interstellar medium can be seen in the spectra of our uncontaminated sample, but unlike pollution by a companion, the interstellar lines are not redshifted at the cluster's velocity ($\sim$232 km s$^{-1}$, \citealt{har96}).

The spectra were obtained in March 2011 and April
2013 using the MXU mode of FORS2 mounted at the VLT on Cerro Paranal,
Chile. Each spectrum is based on the combination of two 2750 s exposures
obtained with the 600B grating and a slit width of 0.7$\arcsec$, and has
a wavelength resolution of $\sim$2.6 \AA. The nominal wavelength
coverage of the sample is 3400-6100 \AA, however some of the spectra
are cut at one end due to their position on the CCD. The spectra were
reduced using a combination of the FORS pipeline (bias subtraction,
flatfielding, wavelength calibration) and a customised IRAF procedure
(extraction, cosmic ray removal, flux calibration).    

A close examination of the spectra revealed that the majority of them
show spectral features of carbon and are at the same time
He-rich objects. The series of plots from Figure \ref{spec}a to Figure \ref{spec}h
illustrates all the spectra in the wavelength range where the
carbon features are most prominent (when present). Note that the
quality of the data is rather remarkable given the relative faintness of
the target stars (they are characterized by a mean $B$ magnitude of
$\sim$18.6). The spectra depicted appear in the same order as the data
summarized in Table 1, i.e., in order of increasing effective
temperature. Note that the carbon lines reach their maximum strength in Figure \ref{spec}f.

\begin{figure}[b]
\includegraphics[angle=0,scale=0.45]{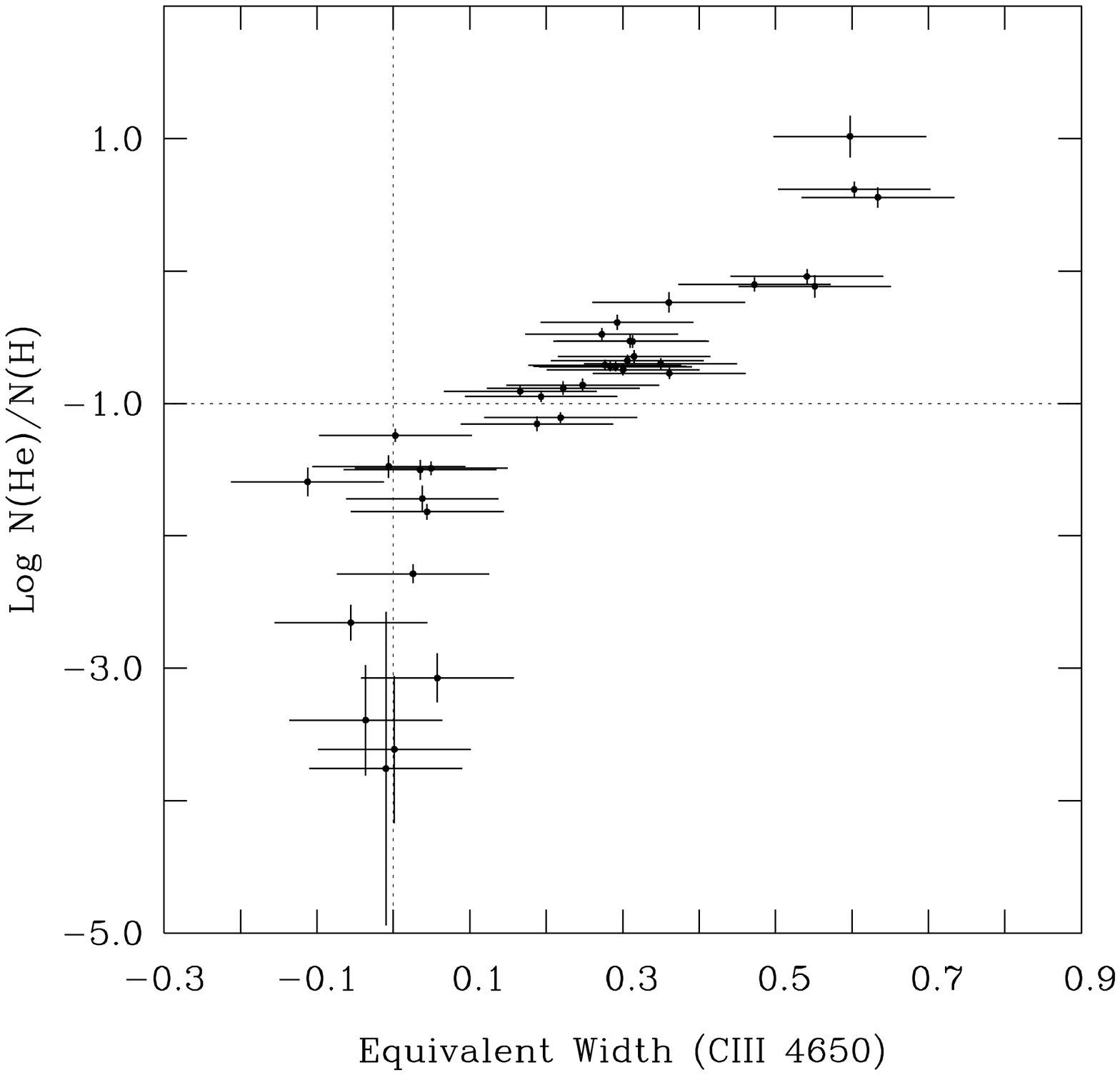}
\caption{ Correlation between the He abundance and the
equivalent width of the CIII 4650 complex detected formally in 25 of our
38 sample stars. The equivalent width is evaluated in arbitrary units.
}
\label{eqw}
\end{figure}

In order to quantify the carbon enhancement, we measured the
equivalent width of the C~\textsc{iii} complex near 4650 \AA\ for each
spectrum and compared this to the derived He abundance as obtained in a
preliminary spectral analysis. The results of this operation are summarized in
Figure \ref{eqw}, which shows that the C~\textsc{iii} feature is detectable in 25 of
our sample of 38 stars. Most interestingly, however, 
Figure \ref{eqw} suggests a
clear correlation between the He abundance and the strength of that
C feature. This finding provided the main incentive to push further and attempt a quantitative
measurement of the carbon abundance through detailed atmosphere modeling.

\section{SPECTROSCOPIC ANALYSIS}

\subsection{Fundamental Parameters}

Given that the spectra appeared to span a
significant range in effective temperature and helium abundance, we decided to 
build dedicated grids of NLTE model atmospheres in order to
 estimate the fundamental parameters of our sample stars in a homogeneous way. Because most
of the stars were expected to be hotter than typical sdBs (He~\textsc{ii}
lines are present in most spectra) and also richer in helium, we
adopted solar abundances for carbon, nitrogen, and oxygen in
our models. This proxy metallicity was adopted simply because these elements are the most important
perturbators of the atmospheric structure of a hot star at their normal
abundances. Note that we did not add iron in our computations because a
solar abundance would not have significantly changed 
the atmospheric structure in the presence of CNO in solar proportions
\citep{haas96,lat11}, and the extra computation time needed outweighed 
the limited benefits. Our model
atmospheres and synthetic spectra were computed with the public 
codes TLUSTY and SYNSPEC and include the following ions (besides those
of H and He) : C~\textsc{ii} to C~\textsc{v}, N~\textsc{ii} to
N~\textsc{vi} and O~\textsc{ii} to O~\textsc{vii}. Note that, as usual
with TLUSTY, the highest ionization stage of each element is taken as a
one-level atom. Additional information on the model atoms can be found on
TLUSTY's Web site\footnote{http://nova.astro.umd.edu/} and in Lanz \&
Hubeny (\citeyear{lanz03,lanz07}). The grid we computed included
models with \teff~between 26,000 K and 58,000 K in steps of 2,000 K, log
$g$ between 5.2 and 6.4 in steps of 0.2 dex, and log~\nhe~from $-$4.0 to
+1.5 in steps of 0.5. That grid was generated by running TLUSTY and
SYNSPEC in parallel mode on our cluster CALYS made up of 320 fast
processors.  

\begin{figure*}[t!]
\includegraphics[scale=0.37,angle=270]{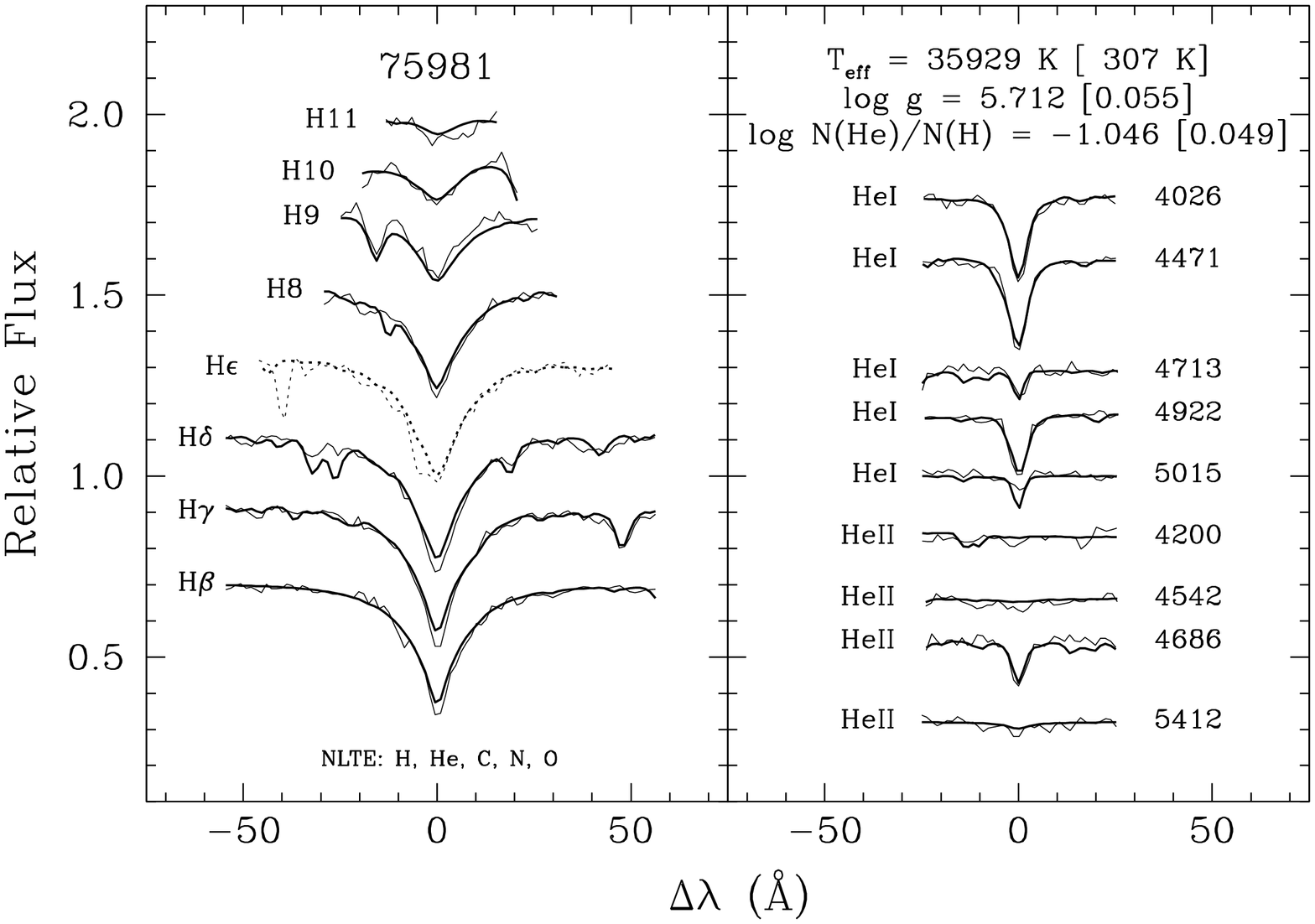}
\includegraphics[scale=0.37,angle=270]{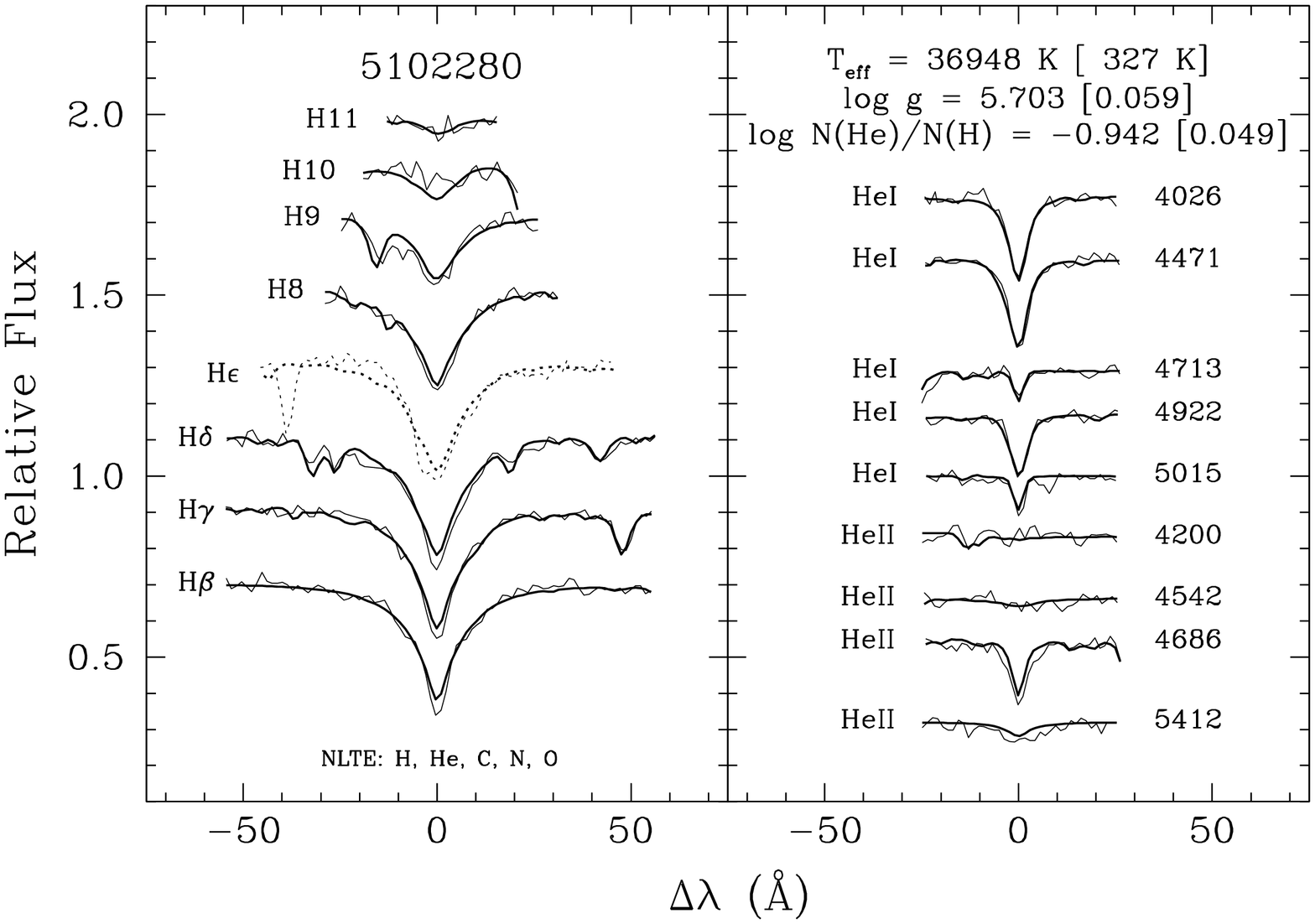}
\includegraphics[scale=0.37,angle=270]{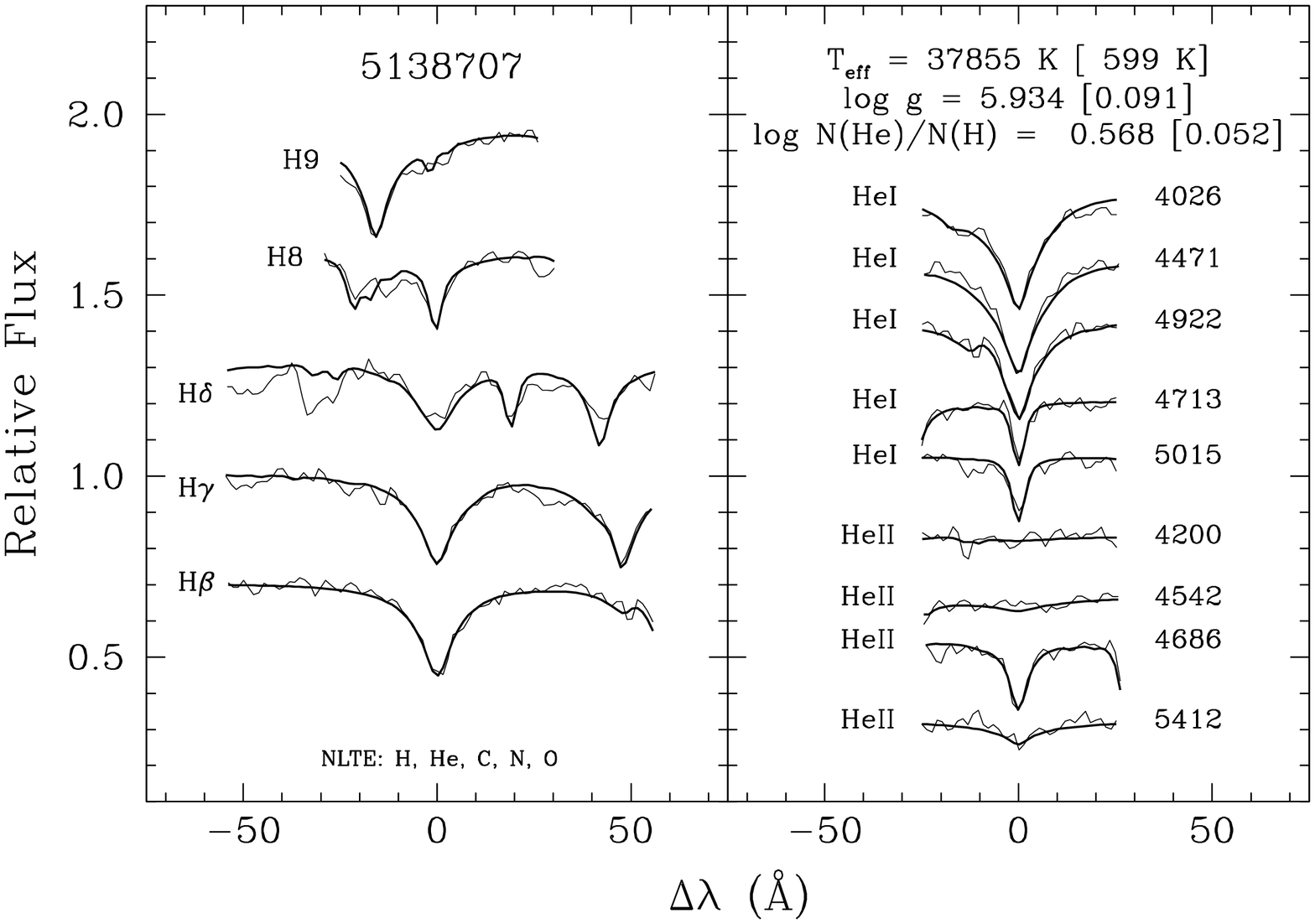}
\includegraphics[scale=0.37,angle=270]{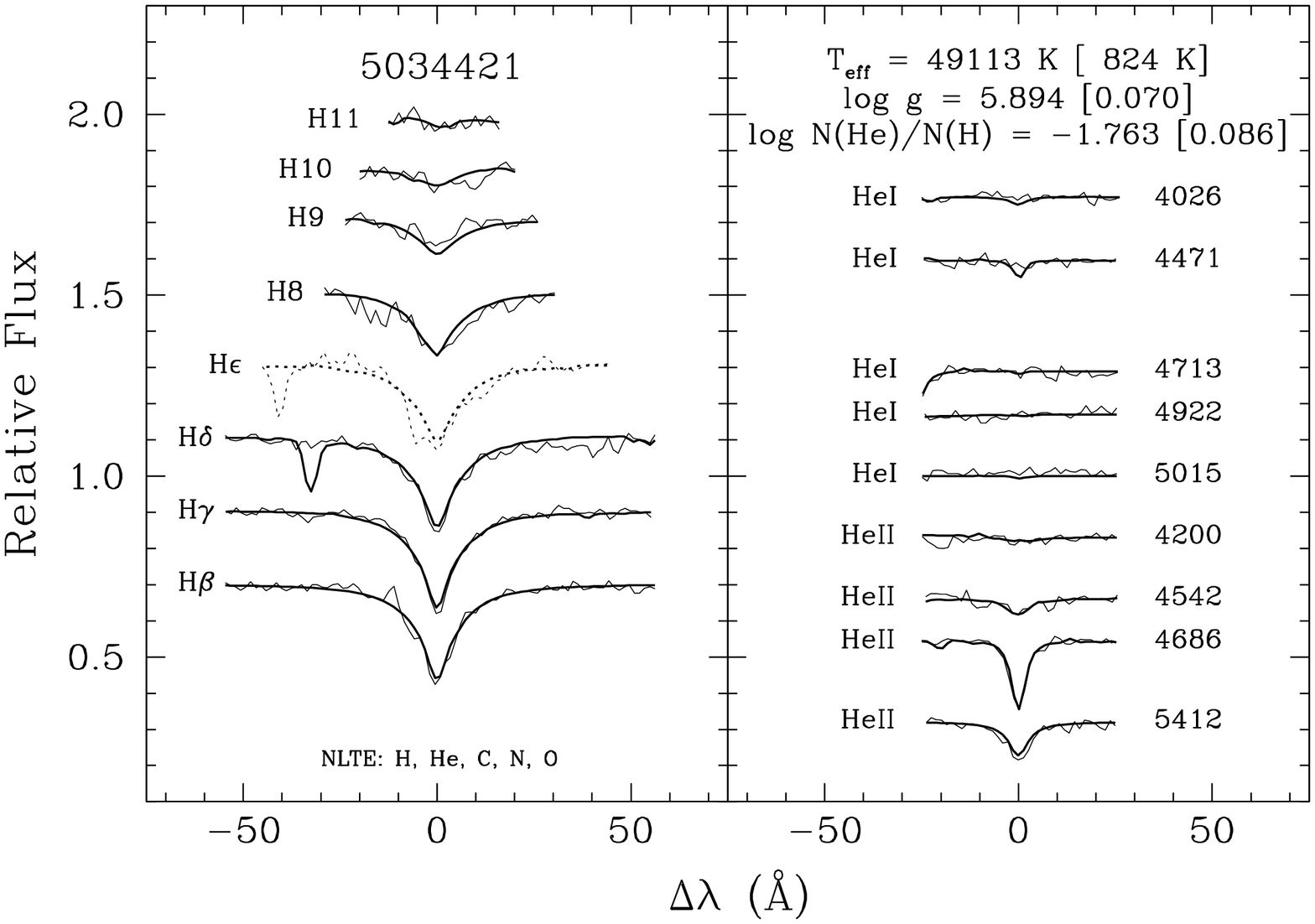}
\caption{Resulting fits for four stars of our
sample.
Top panels (a-b) feature Group 2 stars with a helium
abundance of around solar. The He~\textsc{i} and~\textsc{ii}
lines are well reproduced but some residuals can be seen in the fit to  
the lower hydrogen lines. 
The bottom-left panel (c) shows fit for a helium enriched star, and the
bottom-right (d) one illustrates the fit for a hotter, helium depleted star
\citep[this is the pulsating star identified as V1 in][]{ran11}.Note
that H$\epsilon$, traced by a dashed line, was not included in the fitting
procedure because it is polluted by a fairly strong H component of the
Ca~\textsc{ii} doublet (the K component is also seen in the observed
spectrum).  
}
\label{fit}
\end{figure*}

For most of the stars, the Balmer series (from H$\beta$ up to and
including H11) as well as all the strong helium lines of both ionization
stages (present between $\lambda$5412 in the red and the Balmer jump in
the blue), were simultaneously fit using a $\chi^2$ minimization
procedure similar to that of \citet{saf94}. 
After applying the $\omega$ Cen 
radial velocity shift correction of 232 km s$^{-1}$ to the spectra, the
minimization procedure automatically correct for any residual discrepancies
by matching the core of the observed lines with the modeled ones.
However, a few observed
spectra did not include the He~\textsc{ii} line at 5412 \AA, and a few
others were cut in the blue due to their positions on the CCD chip
so that the higher Balmer lines (between H8 and H11) could not always be
included. In addition, the H$\epsilon$ line was explicitly disregarded
because of the interstellar pollution caused by the H line of Ca~\textsc{ii}.  
Our derived parameters (\teff, log $g$, and log \nhe), as well as the abundances by mass fraction are listed in
Table 1, and Figure \ref{fit}a to Figure \ref{fit}d display representative fits for four
stars. Given the relative faintness of the targets, the results we
achieved are quite satisfactory in terms of simultaneoulsy fitting
all of the available lines. This suggests that the derived atmospheric
parameters are reliable. Note, however, that the quoted
uncertainties refer only to the formal errors of the fits; the true
uncertainties will certainly be larger.

A cursory inspection of our results suggests that our target stars can 
naturally be divided into three groups, and this is indicated in Table 1. 
The seven coolest objects in the table\footnote{5180753 is formally
  hotter than 5142999.} form our Group 1 and are typical
H-rich sdB stars. Our Group 2 is constituted of the 25 following stars,
which are He-rich subdwarfs with log \nhe $\gta$ $-$1.0. Note that for
the present purpose, we will consider 5142999 and 75981 as ``He-rich''
stars even though their helium abundance is slightly below the imposed
limit. Our Group 3 is made of the 6 hottest objects in our sample and is a
collection of hot H-rich sdO subdwarfs, including 4 pulsators (5034421,
177238, 154681, 281063).

\begin{figure}[t]
\includegraphics[angle=0,scale=0.45]{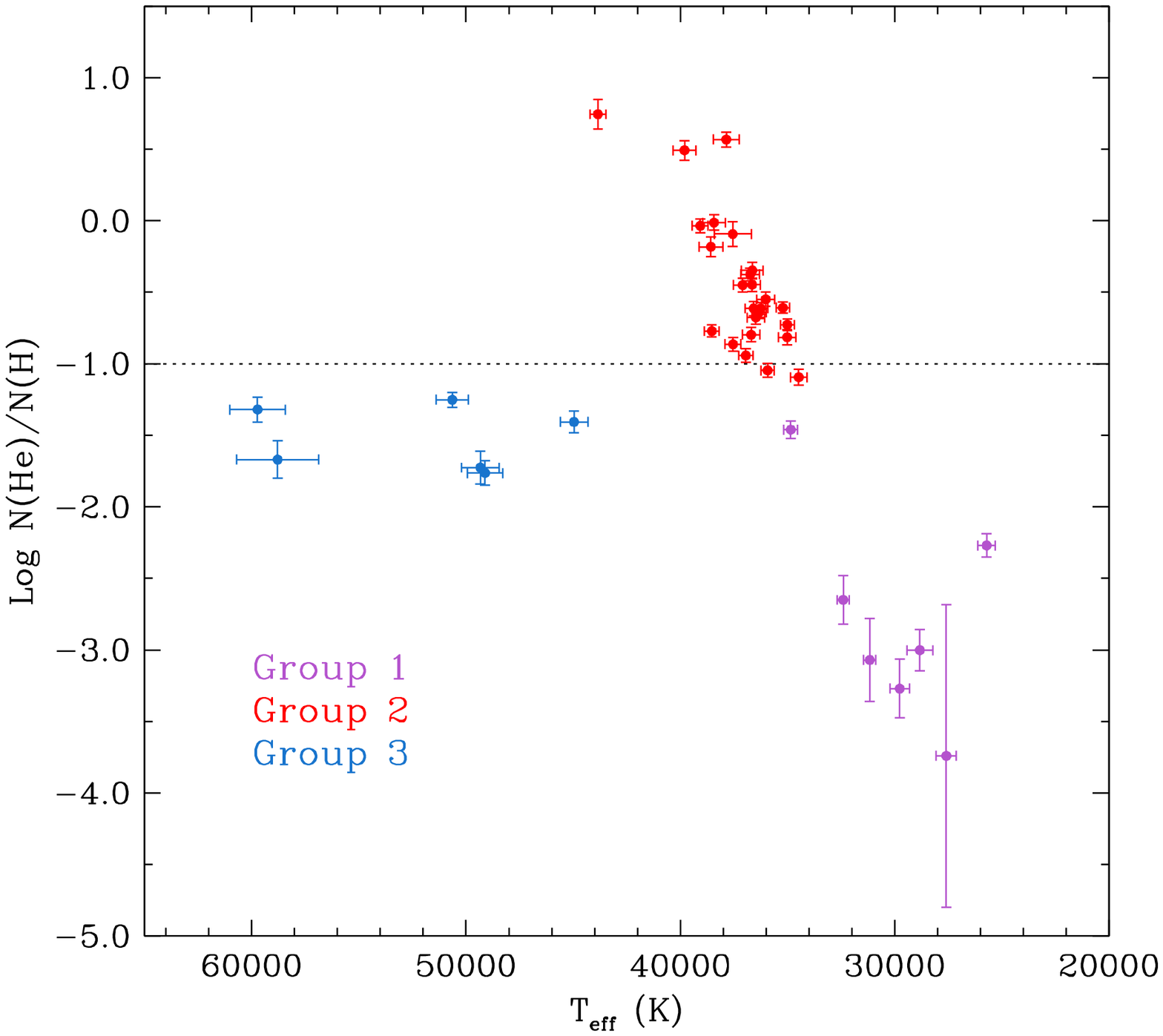}
\caption{Helium abundance versus effective temperature for
the 38 stars of our sample. Group 1 stars are found at lower 
temperatures and are illustrated in purple. Group 2 stars are depicted
in red and are generally He-rich objects. A clear trend of increasing
helium abundance with effective temperature can be noticed among
them. Finally, the hottest stars forming Group 3 are shown in blue. The error
bars include only the formal uncertainties of the fitting procedure and
should be regarded as lower limits. The dotted line indicates the solar
helium abundance.
}
\label{teffhe}
\end{figure}

\begin{figure}[t!]
\includegraphics[angle=0,scale=0.45]{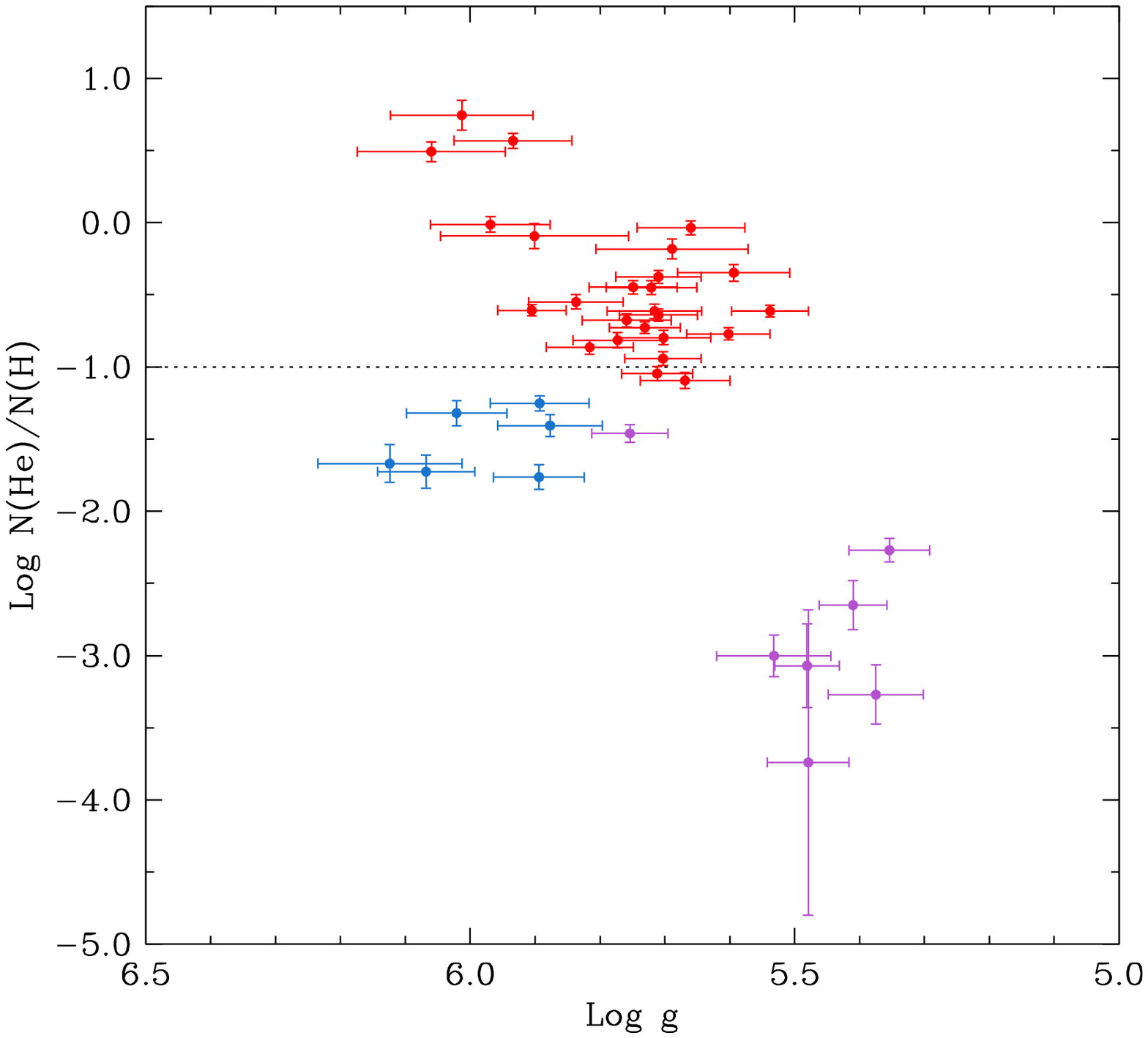}
\caption{Similar to Fig. \ref{teffhe}, but showing helium abundance
versus surface gravity. Though no clear systematic trends are seen, each
group of stars is well-defined in this diagram. The position
of Group 1 stars (in purple) is consistent with typical EHB subdwarf B
stars.  
}
\label{logghe}
\end{figure}

The natural separation between the three groups of stars is easily seen
in Figure \ref{teffhe}, which depicts the helium number abundance (relative to hydrogen)
as a function of effective temperature for the 38 stars of our sample. 
The coolest and hottest stars show significant underabundances of
helium, while among the He-rich objects (Group 2, in red) a positive
correlation is seen between the two parameters. 
This is reminiscent of the relations found by \citet{ede03} for field sdB stars
(see their figure 5).
Figure \ref{logghe} shows our sample in the log \nhe-log $g$ plane, where the six
coolest stars stand out with their low helium abundances
and surface gravities. Combined with the lower effective temperatures inferred,
this implies they are typical helium core burning sdB stars. 
In contrast, the Group 3 stars must certainly be the analogs
of post-EHB H-rich hot sdO subdwarfs on their way to the
white dwarf regime. It is worth mentioning that the Group 3 objects show a He abundance 
lower than the He richest stars of Group 2 by $\sim$2.5 dex, but have
 very similar surface gravities.

%
\begin{deluxetable*}{lccccc}{t}
\tablewidth{0pt}
\tabletypesize{\scriptsize}
\tablecaption{Atmospheric and Other Parameters for the 38 Stars of our Sample } 
\tablehead{ 
\colhead{Number} &
 \colhead{\teff (K)} &
 \colhead{log $g$} & 
 \colhead{log \nhe} &
 \colhead{X(H)} &
 \colhead{X(He)}
 }
\startdata
5238307 & 25711 $\pm$ 400  & 5.35 $\pm$ 0.06  & $-$2.27 $\pm$ 0.08 &
 0.979 $\pm$  0.004 & 0.0209 $\pm$  0.004 \\
5139614 & 27594 $\pm$ 468  & 5.48 $\pm$ 0.06  & $-$3.74 $\pm$ 1.06 &  
 0.999 $\pm$  0.002 &  0.001 $\pm$  0.002\\
204071  & 28828 $\pm$ 602  & 5.53 $\pm$ 0.09  & $-$3.00 $\pm$ 0.14 & 
0.996 $\pm$  0.001 & 0.004 $\pm$  0.001 \\
168035  & 29770 $\pm$ 454  & 5.38 $\pm$ 0.07  & $-$3.27 $\pm$ 0.20 & 
0.998 $\pm$ 0.001 & 0.002 $\pm$ 0.001\\
5262593 & 31161 $\pm$ 280  & 5.48 $\pm$ 0.05  & $-$3.07 $\pm$ 0.29  &
0.997  $\pm$ 0.002 & 0.003 $\pm$  0.002\\
5243164 & 32403 $\pm$ 281  & 5.41 $\pm$ 0.05  & $-$2.65 $\pm$ 0.17  &
0.991 $\pm$  0.003 & 0.009 $\pm$  0.003\\
5180753 & 34850 $\pm$ 317  & 5.75 $\pm$ 0.06  & $-$1.46 $\pm$ 0.06 & 
0.88 $\pm$ 0.02 & 0.12  $\pm$  0.02\\
\hline
5142999 & 34477 $\pm$ 392  & 5.67 $\pm$ 0.07  & $-$1.09 $\pm$ 0.05 &
0.75  $\pm$  0.02 & 0.24  $\pm$  0.02\\
5222459 & 35008 $\pm$ 327  & 5.73 $\pm$ 0.05  & $-$0.73 $\pm$ 0.04  &
0.57  $\pm$  0.02 & 0.42  $\pm$  0.02\\
5119720 & 35018 $\pm$ 403  & 5.77 $\pm$ 0.07  & $-$0.81 $\pm$ 0.05 & 
0.62 $\pm$ 0.03 & 0.38  $\pm$  0.03\\
53945   & 35216 $\pm$ 316  & 5.91 $\pm$ 0.05  & $-$0.61 $\pm$ 0.04 & 
0.50  $\pm$  0.02 & 0.49  $\pm$  0.02\\
75981   & 35929 $\pm$ 307  & 5.71 $\pm$ 0.05  & $-$1.05 $\pm$ 0.05 &
0.74  $\pm$  0.02 & 0.26  $\pm$  0.02\\
5164025 & 36020 $\pm$ 428  & 5.84 $\pm$ 0.07  & $-$0.55 $\pm$ 0.05 &
0.47  $\pm$  0.03 & 0.52  $\pm$  0.03\\
5205350 & 36251 $\pm$ 335  & 5.54 $\pm$ 0.06  & $-$0.61 $\pm$ 0.04 &
0.50   $\pm$ 0.02 & 0.49  $\pm$  0.02\\
5165122 & 36331 $\pm$ 328  & 5.71 $\pm$ 0.06  & $-$0.64 $\pm$ 0.04 &
0.52 $\pm$  0.02 & 0.48  $\pm$  0.02\\
165943  & 36479 $\pm$ 401  & 5.76 $\pm$ 0.07  & $-$0.68 $\pm$ 0.05 & 
0.55 $\pm$ 0.03 & 0.45  $\pm$  0.03\\
5141232 & 36583 $\pm$ 402  & 5.72 $\pm$ 0.07  & $-$0.61 $\pm$ 0.05  &
0.50 $\pm$ 0.03 & 0.49  $\pm$ 0.03\\
274052  & 36640 $\pm$ 506  & 5.59 $\pm$ 0.09  & $-$0.35 $\pm$ 0.06  &
0.36  $\pm$  0.03 & 0.64  $\pm$  0.03\\
5242504 & 36653 $\pm$ 387  & 5.75 $\pm$ 0.07  & $-$0.45 $\pm$ 0.05  &
0.41 $\pm$  0.03 & 0.58  $\pm$  0.03\\
264057  & 36696 $\pm$ 408  & 5.70 $\pm$ 0.07  & $-$0.80 $\pm$ 0.05  &
0.61 $\pm$  0.03 & 0.39 $\pm$  0.03\\
5142638 & 36740 $\pm$ 428  & 5.71 $\pm$ 0.07  & $-$0.38 $\pm$ 0.05  &
0.38 $\pm$ 0.03 & 0.62  $\pm$ 0.03\\
5102280 & 36948 $\pm$ 327  & 5.70 $\pm$ 0.06  & $-$0.94 $\pm$ 0.05 & 
0.68 $\pm$ 0.03 & 0.31  $\pm$  0.03 \\
177711  & 37093 $\pm$ 433  & 5.72 $\pm$ 0.07  & $-$0.45 $\pm$ 0.05 & 
0.41  $\pm$  0.03 & 0.58  $\pm$  0.03\\
5220684 & 37544 $\pm$ 368  & 5.82 $\pm$ 0.07  & $-$0.86 $\pm$ 0.05 & 
0.64  $\pm$  0.03 & 0.35 $\pm$ 0.03\\
5062474 & 37554 $\pm$ 863  & 5.90 $\pm$ 0.14  & $-$0.09 $\pm$ 0.09 &
0.23  $\pm$  0.04 & 0.75  $\pm$ 0.04 \\
5138707 & 37855 $\pm$ 599  & 5.93 $\pm$ 0.09  &  0.57 $\pm$   0.05 & 
0.062 $\pm$  0.007 & 0.92 $\pm$ 0.01\\
5124244 & 38432 $\pm$ 530  & 5.97 $\pm$ 0.09  & $-$0.01 $\pm$ 0.05  &
0.20  $\pm$  0.02 & 0.79  $\pm$  0.02\\
5170422 & 38533 $\pm$ 340  & 5.60 $\pm$ 0.06  & $-$0.77 $\pm$ 0.04  &
0.60 $\pm$  0.02 & 0.40  $\pm$  0.02\\

5047695 & 38578 $\pm$ 549  & 5.69 $\pm$ 0.12  & $-$0.18 $\pm$ 0.07 & 
0.27  $\pm$  0.03 & 0.71  $\pm$  0.03\\
5085696 & 39072 $\pm$ 371  & 5.66 $\pm$ 0.08  & $-$0.04 $\pm$ 0.05 &
0.21  $\pm$  0.02 &  0.78 $\pm$  0.02\\
5039935 & 39804 $\pm$ 523  & 6.06 $\pm$ 0.11  &  0.49 $\pm$ 0.07 & 
0.07  $\pm$  0.01 & 0.91 $\pm$   0.02\\
165237  & 43843 $\pm$ 362  & 6.01 $\pm$ 0.11  &  0.75 $\pm$ 0.10 & 
0.042  $\pm$ 0.001 & 0.95  $\pm$  0.01 \\
\hline
5242616 & 44959 $\pm$ 637  & 5.88 $\pm$ 0.08  & $-$1.41 $\pm$ 0.08 & 
0.86  $\pm$  0.02 & 0.13 $\pm$   0.02 \\
5034421 & 49113 $\pm$ 824  & 5.89 $\pm$ 0.07  & $-$1.76 $\pm$ 0.09 & 
0.92  $\pm$  0.01 & 0.06 $\pm$   0.01\\
177238  & 49328 $\pm$ 877  & 6.07 $\pm$ 0.08  & $-$1.73 $\pm$ 0.11 & 
0.92  $\pm$  0.02 & 0.07 $\pm$   0.02\\
154681  & 50635 $\pm$ 758  & 5.89 $\pm$ 0.08  & $-$1.25 $\pm$ 0.05 & 
0.81   $\pm$ 0.02 & 0.18  $\pm$  0.02\\
281063  & 58789 $\pm$ 1910  & 6.12 $\pm$ 0.11  & $-$1.67 $\pm$ 0.13 &
0.91  $\pm$  0.02 & 0.08  $\pm$  0.02\\
177614  & 59724 $\pm$ 1288  & 6.02 $\pm$ 0.08  & $-$1.32 $\pm$ 0.09 &
0.83  $\pm$  0.03 & 0.16  $\pm$  0.03\\
\enddata
\end{deluxetable*}

\subsection{Helium and Carbon Abundances}

As mentioned earlier, our spectra show a correlation between the presence 
(and strength) of carbon lines and those of helium. This is the same phenomenon
as described by \citet{stro07}, who demonstrated a link between helium enrichment and the
presence of carbon and/or nitrogen lines in field sdO stars. Given that our
spectra of $\omega$ Cen stars are rather limited in resolution and sensitivity,
they are not optimally suited for studying weak metal lines in the
optical domain. Nevertheless, carbon lines were easily found
in our spectra, even for the less helium-rich stars, thanks to the
strong C~\textsc{iii} complexes around 4070 \AA~and  4650 \AA\ 
(see Fig. \ref{spec}). For the weaker nitrogen lines, it was possible to associate
features in the spectra with N~\textsc{ii} and
\textsc{iii} only in the most helium-rich
stars. Therefore, we decided to focus our efforts on the carbon
lines and specifically on quantifying the amount of carbon present in
the atmosphere of these stars. 

In order to accurately derive the carbon abundances, 
we built a small grid of model atmospheres for each star in our sample, keeping 
the fundamental parameters of the models fixed at the values given in Table 1, 
but varying the carbon abundance. Looking back at
the spectroscopic fits we obtained for our Group 1 stars, it became clear
that the solar abundance initially assumed for C, N, and O was far too high, yielding
strong metal lines in the synthetic spectra that were not recovered in the observations. 
In fact, with the exception of a few very weak lines in the hottest star, none of the 
 Group 1 stars show any metal lines whatsoever. Adding to this the annoying tendency 
of our strongest C lines to blend with the O~\textsc{ii} lines (see below), we 
decided to include only carbon as a metal
in our small model grids for Group 1 stars. For the other helium-poor
stars in Group 3, the higher effective temperatures wipe out most of
the metal lines even when included at solar abundance, so we were able to use
 models with the original solar
amount of oxygen and nitrogen. For both of these H-rich groups we varied the carbon
abundance of our models from log \nc = $-$6.0, where no C lines are
visible, to $-$3.5, approximately the solar abundance, in steps of 0.5
dex. 

The stars of Group 2 were subdivided into two categories for the
carbon abundance analysis. Seven of these stars show clear and strong C
lines requiring a super-solar abundance. These seven
stars also happen to be the hottest and most helium-rich members of Group 2.
 Their carbon abundance was obtained with model atmospheres containing 
 nitrogen and oxygen at solar abundance,
while the carbon content was extended up to log \nc = $-$1.0 in the grids.
For the remaining stars in Group 2, which show moderately strong C lines,
the blending of the $\lambda\lambda$4070, 4650 C~\textsc{iii} complexes with the
O~\textsc{ii} lines became problematic, because at the low resolution of our observations 
(2.6~\AA) a solar amount of oxygen in a model without carbon produces small features mimicking weak
C~\textsc{iii} lines. Therefore, we reduced the amount of oxygen to log $N$(O)/$N$(H) = $-$4.5
(roughly 1/10 solar) in the carbon grids for the 18 remaining Group 2
stars. We believe this is a good working assumption, since a careful check of our spectra for
hints of oxygen lines (such as $\lambda\lambda$4415, 4276, 4609 and the
doublet around 4700 \AA) revealed nothing above the noise. 

\begin{figure*}[t]
\includegraphics[angle=270,scale=0.70]{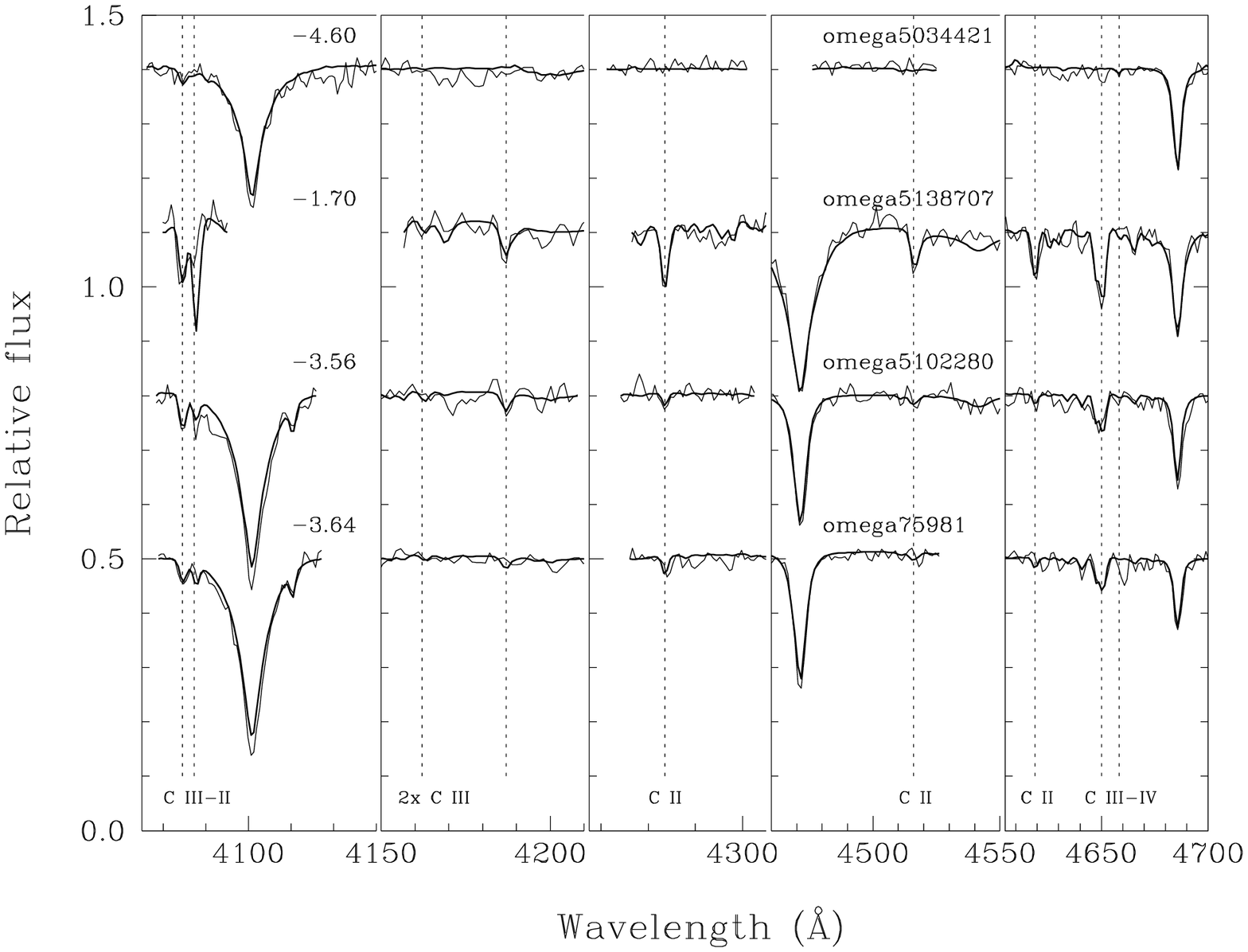}
\caption{Results of our fitting procedure for five
regions containing carbon lines for the four stars presented in
Fig. 3a to Fig. 3d, with their mean log $N$(C)/$N$(H) indicated.
The carbon abundance in the model is that
listed in Table 2 for the respective region. If a region was not
fit, the mean abundance was used in the model. 
}
\label{fitc}
\end{figure*}

Five regions featuring carbon lines were individually fit, including
the C~\textsc{iii} complexes around 4070 \AA~and  4650 \AA,
the C~\textsc{iii} lines at 4162.9 and 4186.9 \AA, the C~\textsc{ii} lines
at 4267.3, 4516.8 and 4619.0 \AA, and C~\textsc{iv} at 4658.3 \AA. The
abundances obtained from these five spectral regions are listed in
Table 2 together with the weighted mean adopted as the final
abundance for each star and the equivalent mass fraction. 
In a few spectra, the C lines were visible only marginally or not at all;
 in these cases we deduced an upper limit on the abundance using the
strongest C line (usually $\lambda$4070). For the remaining stars only the 
clearly visible carbon lines were analyzed, hence the
empty spaces in Table 2. Performing a formal fit on a line was not
always possible, usually because of an uneven continuum, so in these
cases we evaluated the appropriate abundance by eye and assumed an
uncertainty of 0.5 dex. Otherwise, the uncertainties listed
correspond to the formal values derived by the fitting procedure (which is the same as
 used to determine the fundamental parameters but setting only the carbon abundance 
as a free parameter). Examples of our results are plotted in
Figure \ref{fitc} which shows -- for the same four stars whose fits
are displayed in Fig. \ref{fit}a through Fig. \ref{fit}d -- a comparison between the
observed and the model spectrum with the carbon abundance set to that
determined for the displayed spectral region. If a region displayed was not
fit, we used the mean carbon abundance to compute the model spectrum.  We
noticed some systematic differences between the abundances determined from 
the different carbon lines. For instance, the two main C~\textsc{ii} lines
require a C abundance higher than the mean in order to be well 
reproduced (thus they appear too weak in our models), and the C~\textsc{iii}
line at 4070 \AA\ often gives a lower abundance than its counterpart
at 4650 \AA. The C~\textsc{ii} doublet around 4074.5 \AA~also often
appears too strong when using the abundance needed to reproduce the
neighbouring C~\textsc{iii} complex. An example of the latter behaviour
is illustrated for star 5138707 (see Fig. \ref{fitc}). The reason for these systematic 
differences in derived C abundance based on different lines remains unclear.

\begin{deluxetable*}{lccccccl}{h}
\tablewidth{0pt}
\tabletypesize{\scriptsize}
\tablecaption{Inferred Carbon Abundances (log $N$(C)/$N$(H))} 
\tablehead{ 
\colhead{Name} &
 \colhead{C~\textsc{iii} 4070 \AA} &
 \colhead{C~\textsc{iii} 4163, 4187 \AA } & 
 \colhead{C~\textsc{iii} 4650 \AA} &
 \colhead{C~\textsc{ii} 4267 \AA } & 
 \colhead{C~\textsc{ii} 4517 \AA} &
  \colhead{C weighted mean} &
 \colhead{X(C)}
 }
 \startdata 

5238307  & \textless $-$5.00  &      ...     &      ...     &      ...     &      ...     & \textless $-$5.00 & 1.2 $\times 10^{-4}$ \\
5139614  & \textless $-$4.30  &      ...     &      ...     &      ...     &      ...     &\textless  $-$4.30  & 6.0 $\times 10^{-5}$ \\
204071   & \textless $-$5.00  &      ...     &      ...     &      ...     &      ...     &\textless $-$5.00 & 1.2 $\times 10^{-4}$ \\
168035   & \textless $-$5.00  &      ...     &      ...     &      ...     &      ...     &\textless $-$5.00 & 1.2 $\times 10^{-4}$ \\
5262593  & \textless $-$5.00  &      ...     &      ...     &      ...     &      ...     &\textless $-$5.00 & 1.2 $\times 10^{-4}$ \\
5243164  & \textless $-$4.50  &      ...     &      ...     &      ...     &      ...     &\textless $-$4.50 &  3.7 $\times 10^{-4}$ \\
5180753  &\textless  $-$4.50  &      ...     & $-$4.55 $\pm$ 0.70  &      ...     &      ...     & $-$4.52 $\pm$ 0.41 & 3.2 $\times 10^{-4}$\\
\hline
5142999  & $-$3.50 $\pm$ 0.50  &      ...     & $-$3.45 $\pm$ 0.30  &      ...     &      ...     & $-$3.46 $\pm$ 0.26 & 3.1 $\times 10^{-3}$\\
5222459  & $-$4.00 $\pm$ 0.50  & $-$3.70 $\pm$ 0.40  & $-$3.50 $\pm$ 0.20  & $-$4.10 $\pm$ 0.70  &      ...     & $-$3.62 $\pm$ 0.16  & 1.6 $\times 10^{-3}$\\
5119720  & $-$4.00 $\pm$ 0.50  &      ...     & $-$4.70 $\pm$ 1.00  & $-$3.00 $\pm$ 0.80  &      ...     & $-$3.87 $\pm$ 0.39 & 1.0 $\times 10^{-3}$\\
53945    & $-$3.60 $\pm$ 0.50  &      ...     & $-$3.40 $\pm$ 0.20  & $-$3.00 $\pm$ 0.40  & $-$3.00 $\pm$ 0.50  & $-$3.32 $\pm$ 0.16 & 2.8 $\times 10^{-3}$\\
75981    & $-$4.00 $\pm$ 0.50  &      ...     & $-$3.60 $\pm$ 0.20  & $-$2.45 $\pm$ 1.50  &      ...     & $-$3.64 $\pm$ 0.18 & 2.0 $\times 10^{-3}$\\
5164025  & $-$3.30 $\pm$ 0.50  &      ...     & $-$3.30 $\pm$ 0.20  & $-$2.90 $\pm$ 0.60  & $-$2.90 $\pm$ 0.50  & $-$3.22 $\pm$ 0.17 & 3.4 $\times 10^{-3}$\\
5205350  & $-$3.30 $\pm$ 0.50  & $-$3.50 $\pm$ 0.50  & $-$3.50 $\pm$ 0.20  & $-$2.80 $\pm$ 0.50  &      ...     & $-$3.40 $\pm$ 0.16 & 2.4 $\times 10^{-3}$\\
5165122  & $-$4.00 $\pm$ 0.50  &      ...     & $-$4.00 $\pm$ 0.30  & $-$3.00 $\pm$ 0.70  &      ...     & $-$3.88 $\pm$ 0.24 & 8.2 $\times 10^{-4}$ \\
165943   & $-$4.20 $\pm$ 0.50  &      ...     & $-$3.80 $\pm$ 0.50  &      ...     &      ...     & $-$4.00 $\pm$ 0.35 & 6.5 $\times 10^{-4}$\\
5141232  & $-$3.60 $\pm$ 0.50  & $-$3.80 $\pm$ 0.60  & $-$3.60 $\pm$ 0.30  & $-$2.60 $\pm$ 0.30  &      ...     & $-$3.24 $\pm$ 0.19 & 3.4 $\times 10^{-3}$\\
274052   & $-$3.70 $\pm$ 0.50  &      ...     & $-$3.46 $\pm$ 0.30  &      ...     &      ...     & $-$3.52 $\pm$ 0.26 & 1.3 $\times 10^{-3}$\\
5242504  & $-$3.30 $\pm$ 0.50  & $-$3.50 $\pm$ 0.50  & $-$3.40 $\pm$ 0.20  & $-$2.90 $\pm$ 0.70  &      ...     & $-$3.37 $\pm$ 0.17 & 2.1 $\times 10^{-3}$\\
264057   & $-$4.00 $\pm$ 0.50  & $-$3.30 $\pm$ 0.50  & $-$4.30 $\pm$ 0.40  &      ...     &      ...     & $-$3.94 $\pm$ 0.26 & 8.4 $\times 10^{-4}$\\
5142638  & $-$4.00 $\pm$ 0.50  &      ...     & $-$3.50 $\pm$ 0.20  & $-$3.80 $\pm$ 0.90  &      ...     & $-$3.58 $\pm$ 0.18 & 1.2 $\times 10^{-3}$\\
5102280  & $-$3.80 $\pm$ 0.30  & $-$3.50 $\pm$ 0.50  & $-$3.50 $\pm$ 0.20  & $-$3.00 $\pm$ 0.70  & $-$3.60 $\pm$ 0.70  & $-$3.56 $\pm$ 0.15 & 2.2 $\times 10^{-3}$\\
177711   & $-$3.70 $\pm$ 0.50  & $-$3.50 $\pm$ 0.50  & $-$3.00 $\pm$ 0.20  &      ...     & $-$3.40 $\pm$ 0.70  & $-$3.16 $\pm$ 0.17 & 3.4 $\times 10^{-3}$\\
5220684  & $-$4.10 $\pm$ 0.30  &      ...     & $-$4.10 $\pm$ 0.30  & $-$3.10 $\pm$ 0.60  &      ...     & $-$3.99 $\pm$ 0.20 & 7.8 $\times 10^{-4}$\\
5062474  & $-$2.80 $\pm$ 0.50  & $-$2.80 $\pm$ 0.50  & $-$2.50 $\pm$ 0.30  & $-$1.50 $\pm$ 0.50  &      ...     & $-$2.43 $\pm$ 0.21 & 1.0 $\times 10^{-2}$\\
5138707  & $-$2.30 $\pm$ 0.50  & $-$2.30 $\pm$ 0.50  & $-$1.80 $\pm$ 0.10  & $-$1.10 $\pm$ 0.20  & $-$1.70 $\pm$ 0.30  & $-$1.70 $\pm$ 0.08 & 1.5 $\times 10^{-2}$\\
5124244  & $-$2.80 $\pm$ 0.50  & $-$2.80 $\pm$ 0.50  & $-$2.70 $\pm$ 0.20  & $-$1.90 $\pm$ 0.40  & $-$2.30 $\pm$ 0.50  & $-$2.57 $\pm$ 0.15 & 6.5 $\times 10^{-3}$\\
5170422  & $-$4.00 $\pm$ 0.30  & $-$3.50 $\pm$ 0.40  & $-$3.65 $\pm$ 0.20  &      ...     & $-$3.20 $\pm$ 0.40  & $-$3.65 $\pm$ 0.14 & 1.6 $\times 10^{-3}$\\
5047695  & $-$3.20 $\pm$ 0.30  & $-$3.20 $\pm$ 0.30  & $-$3.30 $\pm$ 0.40  & $-$2.30 $\pm$ 0.10  & $-$2.50 $\pm$ 0.50  & $-$2.50 $\pm$ 0.09 & 1.0 $\times 10^{-2}$\\
5085696  & $-$3.10 $\pm$ 0.20  & $-$3.10 $\pm$ 0.50  & $-$2.60 $\pm$ 0.10  & $-$2.00 $\pm$ 0.50  & $-$2.60 $\pm$ 0.10  & $-$2.65 $\pm$ 0.07 & 5.7 $\times 10^{-3}$\\
5039935  & $-$2.40 $\pm$ 0.50  & $-$2.40 $\pm$ 0.50  & $-$1.80 $\pm$ 0.20  & $-$1.04 $\pm$ 0.40  & $-$1.90 $\pm$ 0.50  & $-$1.81 $\pm$ 0.15 & 1.4 $\times 10^{-2}$\\
165237   & $-$1.90 $\pm$ 0.10  & $-$1.70 $\pm$ 0.20  & $-$1.40 $\pm$ 0.10  & $-$1.40 $\pm$ 0.50  & $-$1.00 $\pm$ 0.50  & $-$1.64 $\pm$ 0.07 & 1.2 $\times 10^{-2}$\\
\hline
5242616  & $-$4.40 $\pm$ 0.30  & $-$3.50 $\pm$ 0.50  & $-$4.10 $\pm$ 0.50  &      ...     &      ...     & $-$4.15 $\pm$ 0.23 & 7.2 $\times 10^{-4}$\\
5034421  & $-$4.60 $\pm$ 0.50  &      ...     & $-$4.60 $\pm$ 0.50  &      ...     &      ...     & $-$4.60 $\pm$ 0.35 & 2.7 $\times 10^{-4}$\\\
177238   & $-$4.90 $\pm$ 0.40  &      ...     & $-$4.37 $\pm$ 0.60  &      ...     &      ...     & $-$4.74 $\pm$ 0.33 & 2.0 $\times 10^{-4}$\\\
154681   &      ...     &      ...     & $-$4.60 $\pm$ 0.10  &      ...     &      ...     & $-$4.60 $\pm$ 0.10 & 2.4 $\times 10^{-4}$\\\
281063   &\textless $-$4.50 &      ...     &      ...     &      ...     &      ...     & \textless $-$4.50 & 3.4 $\times 10^{-4}$\\\
177614   &      ...     &      ...     & \textless $-$4.50 &      ...     &      ...     & \textless $-$4.50 & 3.1 $\times 10^{-4}$\\\
\enddata
\end{deluxetable*}

\begin{figure}[t]
\includegraphics[angle=0,scale=0.45]{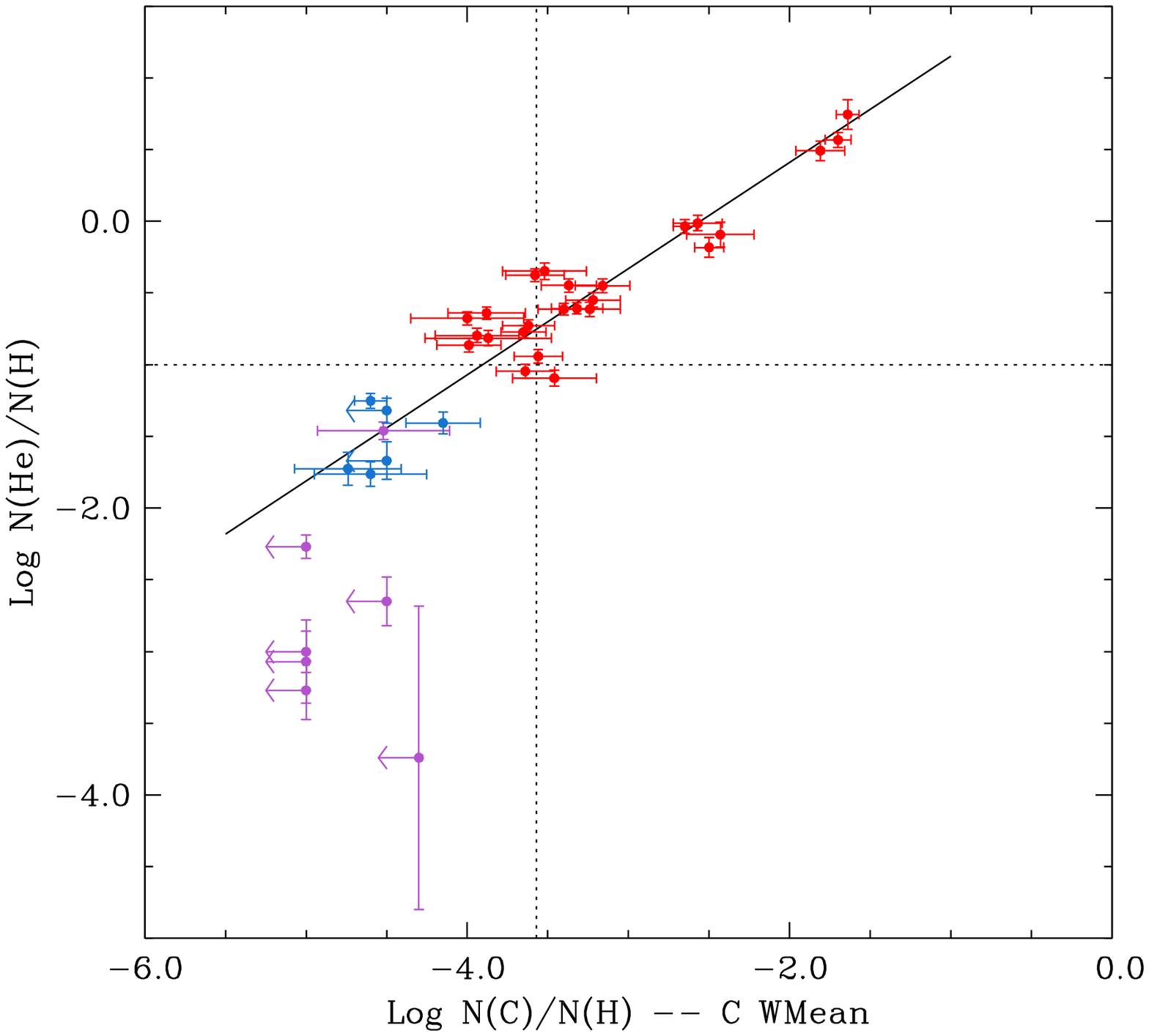}
\caption{ Helium abundance versus the mean carbon
abundance, still using the same color coding as in Fig \ref{teffhe}. The upper limits
on the carbon abundance inferred for eight stars are indicated by arrows
instead of error bars. This diagram shows an obvious relation between the
abundances of the two elements, which is illustrated by the linear
regression (black solid line) based on the 30 stars for which a carbon
abundance was obtained. Dotted lines indicate the solar helium and
carbon abundances. 
}
\label{ncnhe}
\end{figure}

Figure \ref{ncnhe} highlights one of the main results of our
spectroscopic study of EHB stars in $\omega$ Cen: we find a strong correlation
between the helium and carbon abundances. The different groups of stars
are also easily distinguishable in this plot: the most He-rich stars of
Group 2 form the carbon-enriched population, the rest of Group 2 forms a
clump around the 
solar abundance of carbon, and the hot stars of Group 3 are found at the lower
abundance end of the correlation. Finally, for the six coolest Group 1 stars
we can place only an upper limit on the carbon abundance. The Group 1 star 5180753 
is a bit peculiar in that it falls into the low abundance region populated by the Group 3 stars. 
Indeed, it is the only Group 1 star with a measured
carbon abundance. In terms of its (low) effective temperature, this star belongs to Group 1, 
but its carbon and helium abundances, and also to a lesser extend its surface gravity, are more similar to the Group 3 stars. 
This explains its odd position in some of the previous plots.
 It is worth mentioning that in spite of the differences in derived C abundance depending on the line used, the same
correlation is seen for each of the five lines fit. This correlation can be described by a linear
regression between the mean abundance of carbon and the helium abundance as evaluated from the 30 stars 
for which carbon was detected. It is shown in Figure \ref{ncnhe} by the straight black line and can be described
with the equation
\begin{equation}
\begin{split}
{\rm log~N}(\rm{C})/N(\rm{H})=1.36(\pm.039)\times
{\rm log}~N(\rm{He})/N(\rm{H}) \\
 -2.56(\pm.027). 
\end{split}
\end{equation}
A similar positive correlation was found by \citet{nem12} in their study
of field subdwarfs, however their slope is steeper.

We mentioned at the beginning of this section that the nitrogen lines are relatively weak in our spectra and are only discernible in the most helium rich stars. 
Nevertheless, we attempted to roughly estimate the amount of nitrogen present. For the seven most carbon rich stars, we computed model atmospheres with nitrogen abundances of 10 and 100 times the solar value (the original models computed for fitting these stars had a solar abundance of nitrogen) and compared, by eye, the resulting synthetic spectra with those observed. The only useful line for this exercise turned out to be N~\textsc{iii} at 4640 \AA. We estimate that two out of these seven stars may have nitrogen abundances as high as 100 times solar. These two stars are 165237 and 5039935, which are also the most helium and carbon enriched. For the remaining five carbon rich stars the abundances seem to vary between 10 and 100 times solar. For the rest of Group 2, no nitrogen lines are detected above the noise, allowing us to place an upper abundance limit of roughly solar, since this produces lines comparable to the noise level of the observations. So, while the quality and resolution of our spectra 
prevent us from formally fitting and deriving firm abundances for the nitrogen lines, our inspection indicates that N enrichment appears to go hand in hand with helium and carbon enrichment.
\\
\\

\section{DISCUSSION}
\subsection{Comparison with field sdB-sdO stars}

\begin{figure*}[t!]
\includegraphics[angle=180,scale=0.46]{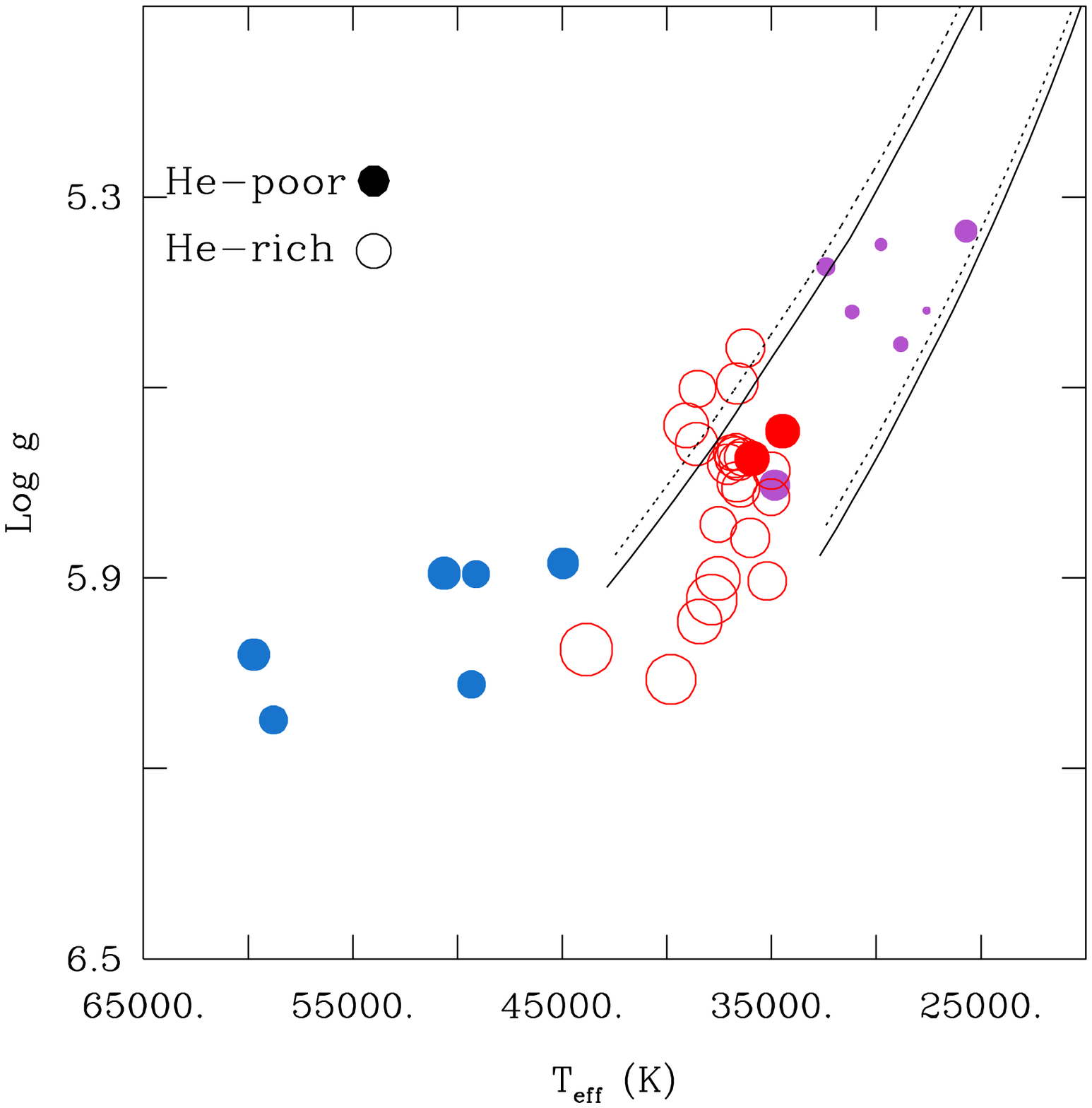}
\includegraphics[angle=180,scale=0.46]{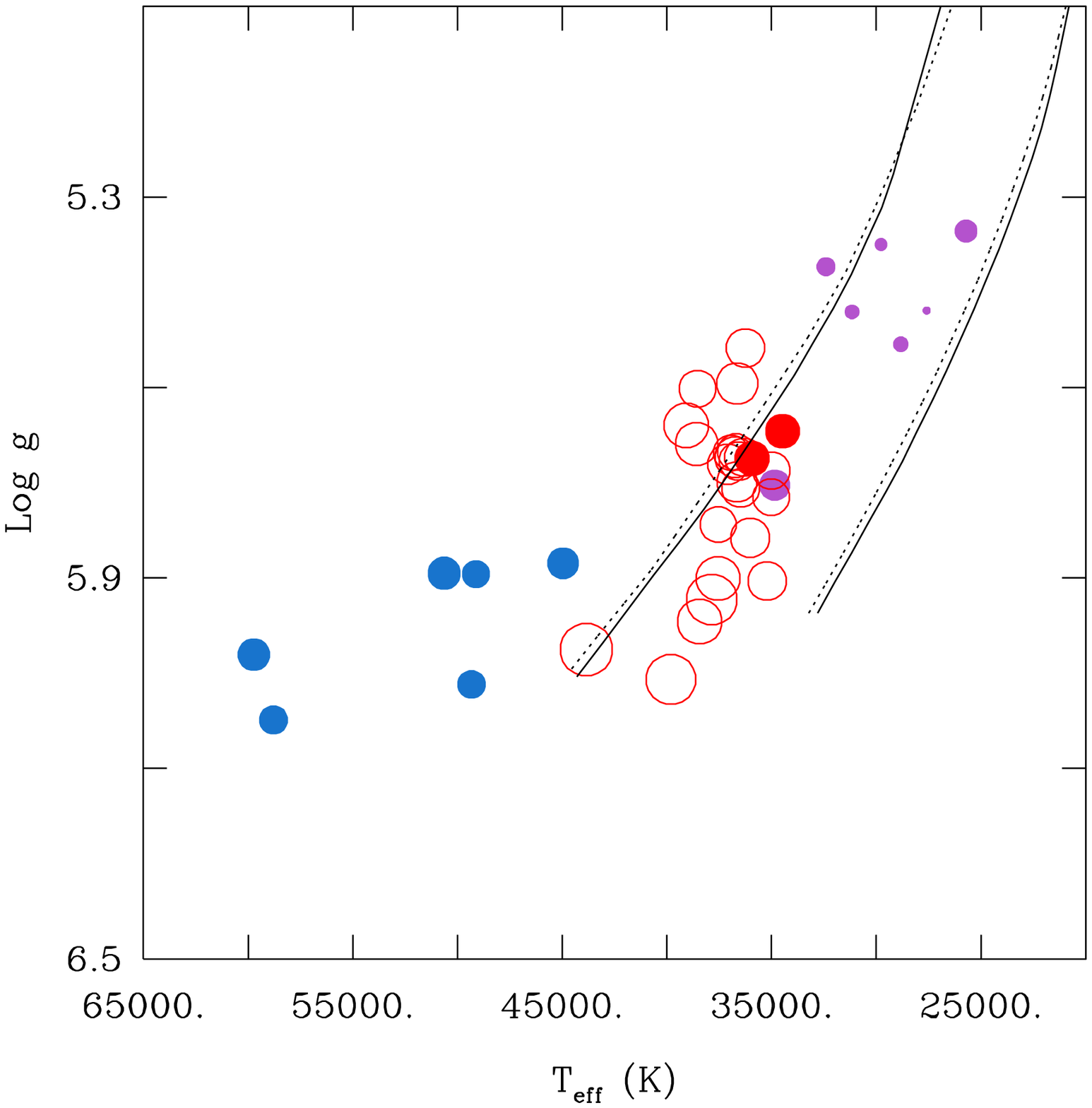}
\caption{Distribution of our sample of $\omega$ Cen EHB
stars in the log $g$-\teff~plane. The size of a given circle is a logarithmic measure of the He abundance relative to that of H. He-poor and He-rich stars are represented by filled and open circles respectively. Left - the ZAEHB and TAEHB are plotted for two different metallicities, Z=0.002 (solid line) and Z=0.0003 (dashed line), and a normal
helium content, Y$\sim$0.24 \citep{pie06}. Right - Here, the EHB was calculated for He-enhanced models (Y=0.4) including metallicities of Z=0.0016 (solid line) and Z=0.0002 (dashed line) (Castellani et al. 2014, in preparation).  
}
\label{loggteff}
\end{figure*}

The distribution of our sample of stars in the log $g$-\teff~diagram is
illustrated in Figure \ref{loggteff}. Here, the size of each circle is proportional to the logarithmic helium abundance,
 filled circles denoting He-poor stars (log \nhe $\lta -$1.0) and open circles indicating He-rich stars.
We also plotted the location of the zero age extreme horizontal branch (ZAEHB) and that of the terminal age extreme horizontal branch (TAEHB) for models with metallicities appropriate for $\omega$ Cen (Z=0.002, solid lines, and Z=0.0003, dashed lines) and normal helium content (Y $\sim$ 0.24) in the left panel \citep{pie06}.
The right panel shows the equivalent tracks for He-enriched models (Y = 0.4) with similar metallicities (Castellani et al., in preparation).
The fact that the coolest H-rich objects (Group 1) are found within the band between the ZAEHB and the TAEHB is consistent with the notion that they are normal helium
core burning sdB stars, while the location of the six hottest
H-rich stars (Group 3) indicates that, as expected, these are
evolved, post-EHB sdO stars. However, the tight clustering of the He-rich objects in our sample (Group 2) around $T_{\rm eff}\sim$40,000 (i.e. on the predicted EHB) is both unexpected and extremely interesting.

\begin{figure}[hb!]
\includegraphics[angle=0,scale=0.46]{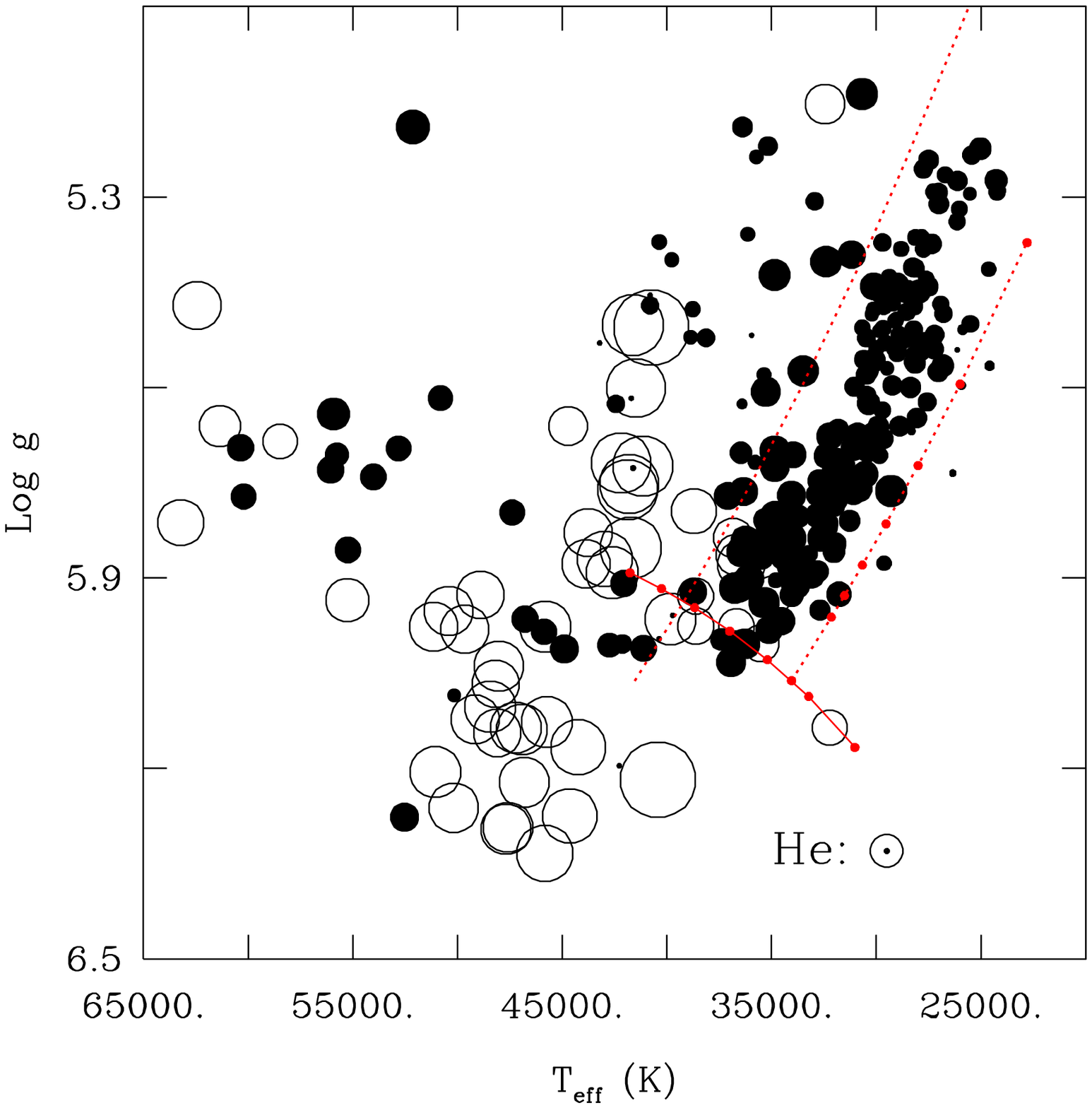}
\caption{ Similar to Fig. \ref{loggteff}, but depicting the
distribution of hot subdwarfs in the field \citep{fon14}. The ZAHEMS, as well as ZAEHB and TAEHB (assuming a core mass of 0.47 \Msun) now refer to models with a metallicity of $Z$=0.02.
}
\label{loggtefffield}
\end{figure}

To put our results for the $\omega$ Cen sample into perspective, Figure \ref{loggtefffield} shows an
equivalent picture for hot subdwarfs in the Galactic field. These data are the result of many years of efforts first
presented by \citet{gre08} through the Arizona-Montr\'eal Spectroscopic
Program and updated recently by \citet{fon14} on the basis of model atmospheres
comparable to those used by us. 
In Figure \ref{loggtefffield}, the ZAEHB, TAEHB and ZAHEMS were computed with our local hot subdwarf evolutionary code at Universit\'e de Montr\'eal. Here, the models have
a metallicity of $Z$ = 0.02, representative of the field stars depicted.
Note, in particular,
how most of the H-rich objects fall in between the ZAEHB and TAEHB
as expected, while the other H-rich stars can be interpreted as
evolved, post-EHB objects on their way to the white dwarf cooling
domain. This is similar to what we find for $\omega$ Cen, with the slight difference that
 the field H-rich sdOs are predominantly found at lower
log $g$ than those in our sample.
But it is the location of the bulk of He-rich stars in the field
in relation to their location in $\omega$ Cen that is particularly
intriguing. While the He-rich stars in the 
field cluster around 45,000 K and are clearly hotter than the He-core burning EHB, the majority of their cluster counterparts are distinctly cooler and in line with the predicted EHB.
Note that the field distribution of He-sdOs as described in the independent studies of \citet{stro07} and \citet{nem12} is very similar to the results shown in Figure \ref{loggtefffield}. 

In view of our findings we checked the independent spectroscopic
study of \citet{moe11} which, among other things, led to the 
characterization of 17 He-rich subdwarfs in $\omega$ Cen\footnote{Note that the majority
of their helium-rich subsample consists of different stars from ours; only three stars are present in both samples: 53945, 75981, and 5142999 (164808 in \citealt{moe11})}. A look at their Table 4 shows
that they also infer effective temperatures around or below 40,000 K for most of
the He-rich stars. In fact, their helium rich subsample is 
even found at slightly lower temperatures than ours.  
It is not clear if this small difference is a
systematic effect related to the way in which the atmospheric parameters were estimated or the consequence of a different color cutoff in the sample favouring hotter stars for our study. 
Interestingly, the sample of EHB stars analysed in NGC 2808 by \citet{moe04} comprises a modest number of He-enriched stars that are found at similar temperatures as those in $\omega$ Cen.

The difference in temperature between He-rich EHB stars in $\omega$ Cen and the field was
 already reported in a review by \citet{heb09}, together with the finding that 
the field population contains a higher fraction of stars with
strongly enriched helium abundances. Indeed, only a few stars in our
and Moehler et al.'s sample have log \nhe\ \textgreater\ 0.0, whereas
such highly He-enriched stars are the rule in the field He-sdO samples of 
\citet{stro07}, \citet{nem12}, and \citet{fon14}. Only a few stars in 
these three samples have parameters comparable to the helium-rich population in 
$\omega$ Cen. These differences point towards fundamental differences between the 
helium-enriched EHB star population in the field and in $\omega$ Cen, and are likely 
related to the fact that sdB and sdO stars in GCs have older (12$-$13 Gyr) and typically 
metal poorer progenitors than their field counterparts. 

The hot subdwarf stars found in the Sloan Digital Sky Survey (SDSS) 
might shed some light on these differences. \citet{hirschthesis} carried out
a spectral analysis of a sample of hot subdwarf spectra selected from the SDSS catalog (Data Release 7) and found a population of stars quite similar to those in $\omega$ Cen in terms of
the \teff\ $-$ log \nhe\ distribution (see his Figure 7.1 and also \citealt{heb06}). These stars have helium abundances between solar and log \nhe = 0.0, and effective temperatures mostly below 40,000 K. 
A preliminary spectral analysis of hot subdwarfs from SDSS DR10 (P. N\'{e}meth, priv. comm.) reveals a similar population. The space distribution and kinematics of the \citet{hirschthesis} sample suggest that most of the stars belong to the halo population, which is likely similar to that of $\omega$ Cen in that it is relatively old and metal-poor. In contrast, most surveys of bright field stars favour metal-rich population I stars, which might explain why the atmospheric properties derived are different.

\subsection{Comparison with evolutionary theory}

The different characteristics and properties of the stars in our sample
 are likely to bear traces of their evolutionary history. The question is how were
these stars formed? As already indicated in the Introduction, the answer to this is by no means straightforward.

For the helium-poor stars in our sample (Group 1 and 3) the evolutionary status at least is easier to
pin down since their atmospheric parameters are consistent with them being 
typical EHB stars (Group 1) and hotter evolved post-EHB stars (Group
3). Let us remember that canonical evolution identifies EHB stars as the
progeny of red giant stars that are subject to important mass loss
before or at the helium flash, leaving behind a helium core burning star
stripped of most of its hydrogen envelope. 
According to Figure \ref{loggteff}a, our modest sample of hydrogen-rich sdBs sits right on the EHB as predicted for models with a canonical He-abundance (Y$\sim$0.24). A large fraction of our helium-enriched stars (Group 2) also lies between the ZAEHB and the TAEHB, which is not unexpected since most of them should be in a central helium-burning phase. 
Models appropriate for the He-enhanced subpopulation in $\omega$ Cen (Y$\sim$0.4), see Figure \ref{loggteff}b) shift the EHB to lower effective temperatures and place the Group 2 stars very close the TAEHB, where the evolution dramatically speeds up.

As mentioned earlier, there are two competing evolutionary scenarios that predict a He-enhancement in the atmosphere of EHB stars, the He-enhanced scenario \citep{dan10} and the late-helium flash \citep{cas93,bro01}. While both scenarios yield a helium-enriched atmosphere (note though that the He mass fraction is not expected to rise above $\sim$ 0.4 in the He-enriched scenario), there is one fundamental measureable difference: stars formed via the He-enhanced scenario are not expected to show any metal enrichment whereas the mixing (and burning) in a typical hot flasher event should enrich the atmosphere not only with helium but also with carbon, and to a lesser extend, nitrogen \citep{cass03}. 
The relation we find between helium and carbon enrichment (as well as the high nitrogen abundances suspected in the most enriched stars) thus strongly indicates a hot flasher origin for our Group 2 stars. 
At the quantitative level however, the situation is more complicated. The deep mixing occuring during a hot flash is expected to consume most of the hydrogen and leave behind an atmosphere composed of approximately 96\% helium by mass and 3 to 4\% carbon. But in our sample only the three most helium-rich stars have a mass fraction of helium higher
than 90\%, with a maximum carbon mass fraction of 1.5\%. These
three stars could still potentially fit within the framework of a late-flash
event, but the bulk of our Group 2 stars have substantially lower helium abundances and cannot be reproduced by this scenario. 
We should mention that an intermediate type of
flash mixing (known as ``shallow mixing'' as opposed to ``deep mixing'') was proposed and studied by \citet{lanz04} and
\citet{mil08}. During a shallow mixing event, the inner and outer parts of the star are not mixed as efficiently; in particular, 
hydrogen is only diluted (not burned) with the helium and carbon-rich material dredged up from the core.
Therefore, the amount of hydrogen remaining in the envelope is higher in a shallow mixing case. 
While this scenario seems to fit our measured abundances better than deep mixing event, the latter is the more usual outcome of a hot flasher event.

\begin{figure}[t]
\includegraphics[angle=0,scale=0.46]{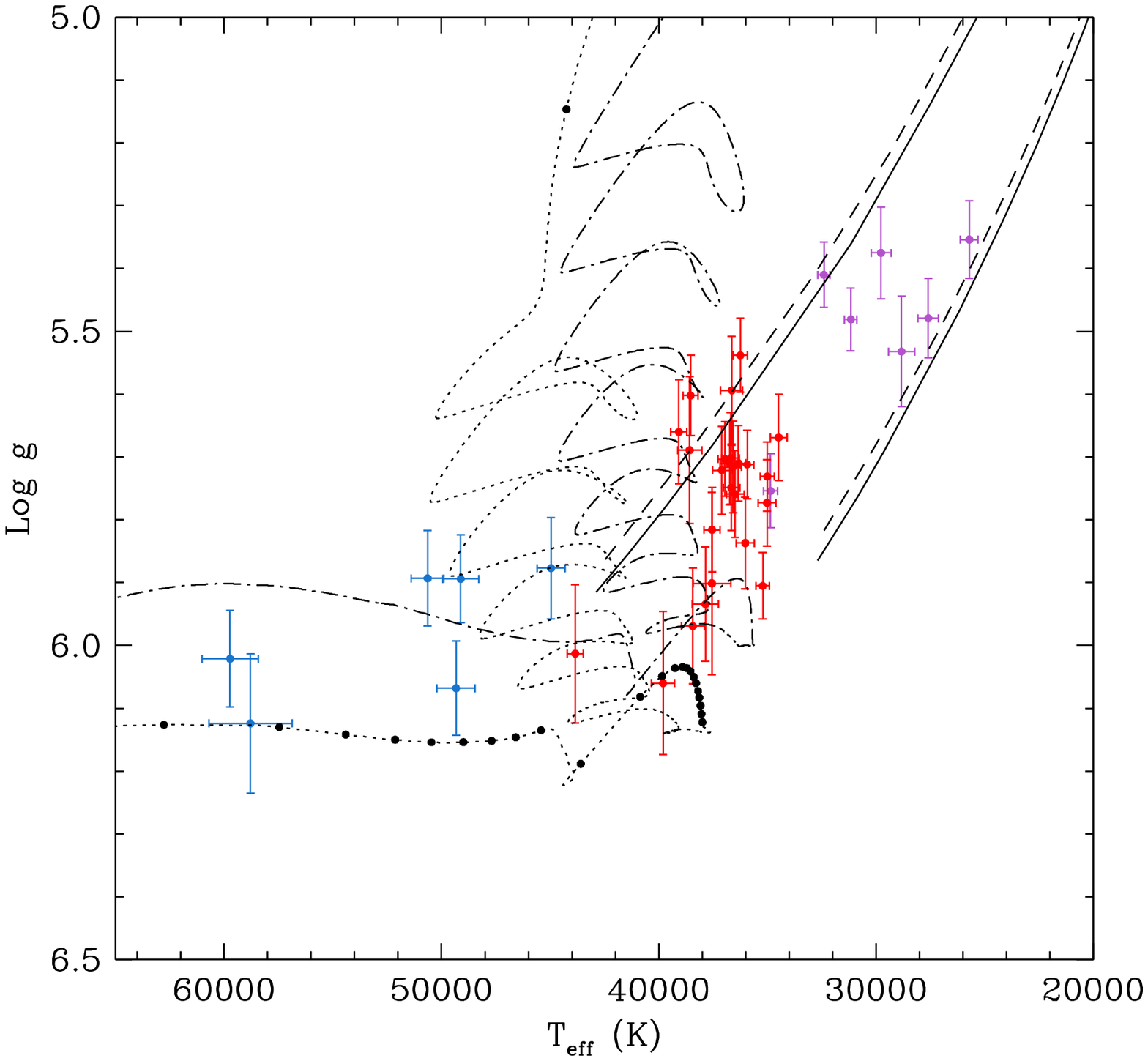}
\caption{ Comparison of our sample with late-flasher evolutionary tracks by \citet{mil08}. The tracks are for a metallicity Z=0.001 and respectively refer to a deep-mixing event (M=0.48150 \Msun, dotted line) and a shallow mixing event (M=0.49145 \Msun, dashed-dotted line). Points at 5 Myr intervals are shown on the first track, to give an idea of the evolutionary timescale in the different regions.
}
\label{lft}
\end{figure}

Examples of late flasher evolutionary tracks \citep{mil08} are shown in Figure \ref{lft} overplotted with our sample. It is obvious that the model tracks do not reproduce the observations very well and in particular predict higher values of log $g$ on the EHB (where the stars spend most of their He-burning life) than measured. This may be a consequence of the fact that neither the shallow nor the deep mixing models consider gravitational settling. Indeed calculations indicate that the inclusion of gravitational settling moves the HB tracks towards lower effective temperatures and surface gravities \citep{moe04,mich11}. Moreover, the late flasher models do not account for the mean molecular barrier when dealing with the convective zones, which may also affect the ZAHB location of the models. 

\begin{figure}[t]
\includegraphics[angle=0,scale=0.46,clip=true]{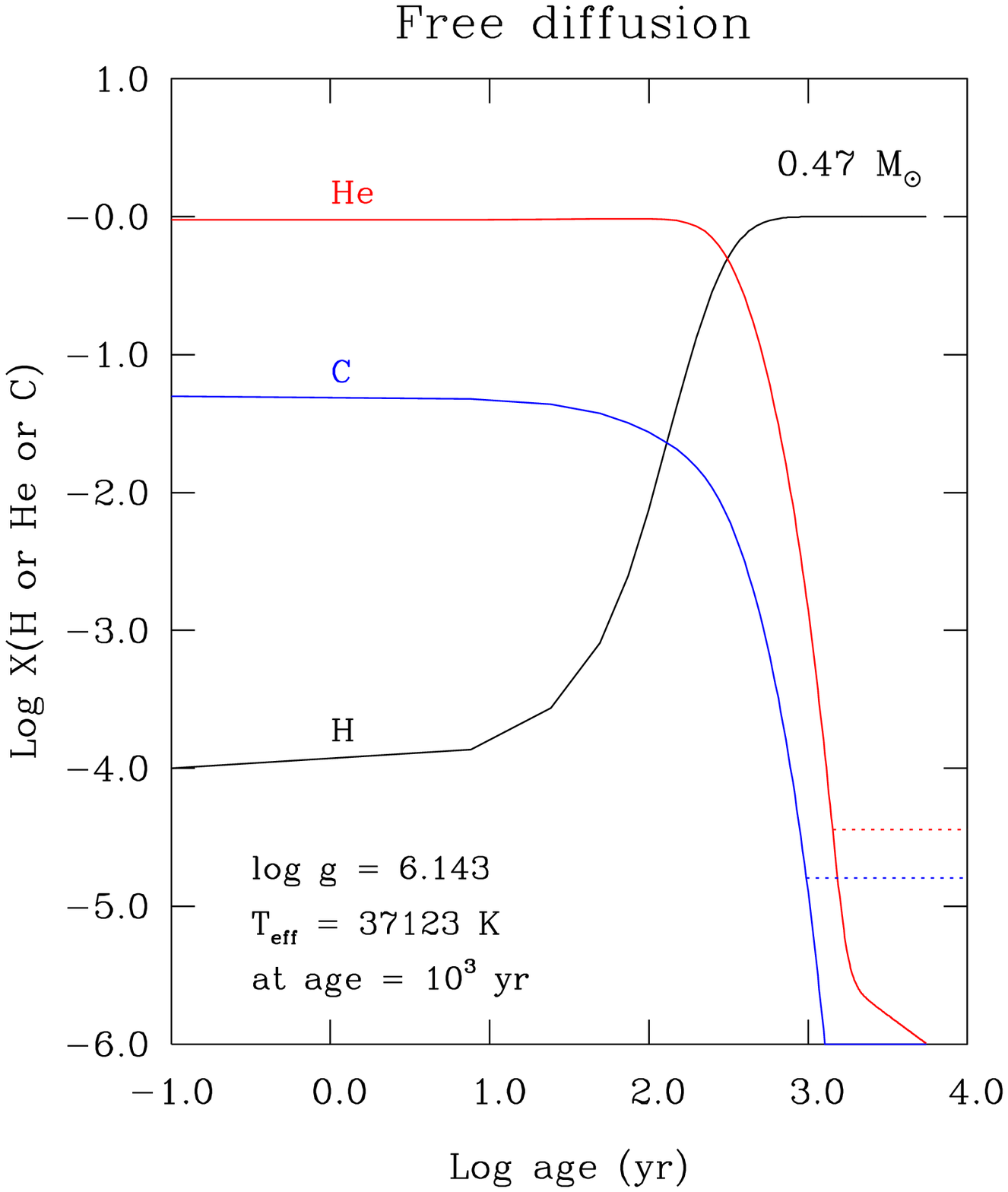}
\caption{ Mass fraction of H, He and C 
as a function of age for an evolutionary model including gravitational settling.
After roughly 10$^3$ yr, the initial helium and carbon rich composition of the atmosphere has become hydrogen dominated. The amount of He and C
that can be supported by radiative levitation is indicated by dotted lines.}
\label{diff}
\end{figure}

When modeling the evolution of stars as compact as those on the EHB it is essential to take into consideration the 
diffusion processes occuring in their atmospheres. Assuming that He-rich subdwarfs are born with more or less the 
atmospheric composition predicted by the late flasher scenario, the heavier elements will quickly dissipate from the atmosphere  
due to gravitational settling if there are no competing mechanisms to slow down this process. 
Indeed, the time needed after the primary helium flash for the star to settle on the ZAEHB is 
of the order of $\sim$10$^6$ yr (\citealt{mil08,bro01}), while diffusion if left unimpeded will 
transform the initially He-rich atmospheric composition to one dominated by H on a timescale of only $\sim$10$^3$ yr.
This is illustrated in Figure \ref{diff}, where we show the evolution of the
surface abundances of H, He, and C (in mass fraction) for a 0.47 \Msun\ subdwarf with an initial atmospheric 
composition as predicted by the late flasher scenario.
Hence, diffusion must be slowed down in the He-rich subdwarfs if they are to be detected as such. The most
likely mechanisms for this are stellar winds or
internal turbulence \citep{hu11}. Note that radiative levitation is not a dominant contributor to the slowing down of gravitational settling,
 since it can only maintain a subsolar amount of carbon and helium 
in the atmosphere of a subdwarf such as the one modeled in Figure \ref{diff}. 

The relation we find between the carbon and the helium abundance
(Figure \ref{ncnhe}) for EHB stars in $\omega$ Cen is also observed in field EHB stars \citep{nem12},
and can largely be interpreted as the signature of diffusion effects. 
The fact that the correlation between the C
and He abundances is positive and that the slope (1.36) is
larger than 1 in a log-log C vs He abundance plot is a
strong indication that chemical separation is going on in these stars,
albeit slowed down by a competing agent, with carbon sinking faster than
helium as can be expected. 
Interestingly, in this plot stars belonging to Group 2 and 3 follow the same relation, thus suggesting a possible evolutionary link between them. The hot sdOs could be post late flashers rapidly evolving towards the white dwarf cooling sequence. \citet{mil08} suggested that due to diffusion effects, the He-rich late flashers could turn into hydrogen-rich objects before approaching the white dwarf regime. A common origin for these two groups of stars could offer an explanation for their relatively high surface gravity. Canonical post-EHB evolutionary tracks usually predict rising luminosities after core He-exhaustion and can account rather well for the lower surface gravities measured for the hotter field sdOs ($\log{g}\lesssim$5.0, cf. Figure \ref{loggtefffield}). However, this increase in luminosity becomes less important for thinner H-envelopes \citep{dor93}. This is why the post-EHB evolution in the hot flasher tracks of Figure \ref{lft} proceeds at relatively constant surface gravities. 


In conclusion, we clearly established the presence of a positive correlation between helium and carbon enhancement in our sample of $\omega$ Cen EHB stars. Such a relation has been suspected for many years (\citealt{moe02}, \citeyear{moe07}, \citeyear{moe11}), but has never been quantified until now. By measuring the carbon abundances in our spectra using dedicated models we were able to confirm that helium enhancement is directly linked with carbon abundances being around or above solar. The linearity of the He-C relation (in the logarithmic plane) likely bears the signature of diffusion effects, but also requires a formation mechanism that enriches the surface of the star not only with helium but also with carbon. Therefore, our results strongly favor a late flasher history over the He-enhanced scenario for our helium-enriched stars. In fact, three quarters of our sample could fit within this framework\footnote{It must be kept in mind though that our sample in strongly biased towards single stars, since 
spectra showing obvious signs of pollution by a companion or nearby star were excluded from our analysis.}. Unfortunately, current evolutionary models for late hot-flashers cannot fully explain the characteristics of our objects; their position in the log $g$-\teff\ diagram is not correctly reproduced and diffusion certainly needs to be taken into account to recover the chemical compositions measured. We are hopeful that the results presented here will trigger the development of more sophisticated models fine-tuned to the EHB star population of $\omega$ Cen and eventually help solve the evolutionary mystery.

\acknowledgments
This work was supported in part by the NSERC Canada through a doctoral
fellowship awarded to M. L. and through a research grant
awarded to G. F. The latter also acknowledges the
contribution of the Canada Research Chair Program.
M. L. also acknowledges funding by the Deutsches Zentrum für Luft- 
und Raumfahrt (grant 50 OR 1315). 
This work was partially supported by PRIN--INAF 2011 "Tracing the
formation and evolution of the Galactic halo with VST" (P.I.: M. 
Marconi)
and by PRIN--MIUR (2010LY5N2T) "Chemical and dynamical evolution of
the Milky Way and Local Group galaxies" (P.I.: F. Matteucci).

 \bibliographystyle{apj}
 \bibliography{referencems}

\end{document}